\documentclass[a4paper,11pt]{article}
\pdfoutput=1 % if your are submitting a pdflatex (i.e. if you have images in pdf, png or jpg format)

\usepackage{jcappub} % for details on the use of the package, please
\usepackage{amsmath}

\usepackage[toc,page]{appendix}
\usepackage{graphicx}
\usepackage{color}
\usepackage{tensor}

\usepackage{xcolor}
\usepackage{graphicx}
\usepackage{colortbl}
\usepackage{placeins}
\usepackage[margin=2.5cm]{geometry}
\usepackage{multirow}

\usepackage[utf8]{inputenc}

\newcommand{\co}{\mathbin{\|}}

\newcommand{\cx}{\mathbin{\times}}

\title{\boldmath Pixel space convolution for cosmic microwave background experiments}

\arxivnumber{}
\author[a]{P. Flux\'a,}
\author[b]{M.K. Brewer,}
\author[a]{R. D\"unner}

\affiliation[a]{Instituto de Astrof\'isica, Pontificia Universidad Cat\'olica de Chile ,\\Vicu\~na Mackenna 4860, Chile}
\affiliation[b]{Department of Astronomy, Johns Hopkins University,\\Baltimore MD, USA}

\emailAdd{pafluxa@astro.puc.cl}
\emailAdd{brewer@astro.umass.edu}
\emailAdd{rdunner@astro.puc.cl}

\abstract{
Cosmic microwave background experiments have experienced an exponential increase in complexity, data size and sensitivity. One of the goals of current and future experiments is to characterize the B-mode power spectrum, which would be considered a strong evidence supporting inflation. The signal associated with inflationary B-modes is very weak, and so a successful detection requires exquisite control over systematic effects, several of which might arise due to the interaction between the electromagnetic properties of the telescope beam, the scanning strategy and the sky model. In this work, we present the Pixel Space COnvolver (PISCO), a new software tool capable of producing mock data streams for a general CMB experiment. PISCO uses a fully polarized representation of the electromagnetic properties of the telescope. PISCO also exploits the massively parallel architecture of Graphic Processing Units to speed-up the main calculation. This work shows the results of applying PISCO in several scenarios, included a realistic simulation of an ongoing experiment, the Cosmology Large Angular Scale Surveyor.}

\begin{document}
\maketitle
\flushbottom

\section{Introduction}

The study of the cosmic microwave background (CMB) in the last few decades has lead to major advances in Cosmology. In particular, the study of the temperature and polarization anisotropy field has allowed us to achieve percent level constraints on cosmological parameters, favoring a universe dominated by cold dark matter plus a cosmological constant ($\Lambda$CDM), known today as the standard model of Cosmology (see \cite{2016A&A...594A..12P}). Current efforts are mostly focused on the polarization anisotropy field, as it promises to provide independent constraints on the very early stages of the Universe. At these very first moments, the leading theory predicts that the Universe exponentially expanded nearly 60 e-folds in a process called inflation, becoming the flat, homogeneous and isotropic Universe that we see today. If true, gravitational waves generated by this expansion would have later interacted with the surface of last scatter, leaving a unique imprint in the polarization field which should be observable today.

Electromagnetic radiation from the CMB anisotropy field is nearly 10\% polarized. These polarized anisotropies can be modeled as a spin-2 field, which is separable in two orthogonal components: a curl-free component called E-mode, and a divergence-free component called B-mode. In the absence of gravitational waves, the polarization field would contain only E-modes, which would be partially turned into B-modes by gravitational lensing. The resulting B-mode field would be one or more orders of magnitude lower than the E-mode signal, being very faint and difficult to measure. Moreover, recent results show that this signal lies below other polarized emissions, like the interstellar medium (see \cite{2018PhRvL.121v1301B}), indicating that detection of the primordial B-mode signal is indeed a major technical challenge.

Another source of confusion is experimental systematic effects. The tremendous increase in sensitivity of CMB experiments has brought into play a long list of instrumental and observational effects, which need to be properly incorporated not only into the data reduction process, but into the design of future experiments as well. Because the CMB is an extended signal on the sky, it is crucial to characterize the beam, or in the more general case, the polarized antenna response of the telescope. It is also key to model and understand how this response gets convolved with the sky signal via the given scanning strategy, as this process may directly affect the scientific results.

Modeling the effect of the telescope beam on scientific results can be done analytically or numerically. For instance, the effect of symmetrical beams with smooth profiles can be analytically accounted for when computing the power spectra of CMB maps (see \cite{2003ApJS..148...39P}). The impact of other related systematic effects, like beam mismatch, has also been described in the literature (see \cite{PhysRevD.77.083003, 2007MNRAS.376.1767O, 2015JCAP...03..048D}). When the beam has more complex features, like sidelobes, numerical methods must be used instead. Several techniques that rely on harmonic space representations of the beam (see \cite{2010ApJS..190..267P}) the sky and the scanning strategy have been devised (see \cite{2001PhRvD..63l3002W,2000PhRvD..62l3002C}). An excellent description of an implementation of this is given in \cite{2018arXiv180905034D}. Another approach is the one followed by FeBeCop (see \cite{2011ApJS..193....5M}), which computes an effective beam instead of convolving the sky for every time sample. 

In this work, we describe the algorithm and prototype implementation of a new CMB computer simulation code, the Pixel Space COnvolver (PISCO). PISCO has the capability of accounting for arbitrary shaped, time-varying beams and sky models. Polarization properties of the beam are handled following the work presented in \cite{2007MNRAS.376.1767O}. PISCO also exploits the massively parallel architecture that Graphics Processing Units (GPU) provide via the CUDA application programming interface, and was designed from scratch to take advantage of HPC environments.

This paper is organized as follows. 
\S\ref{sec::coordinate-systems} contains a description of the coordinate system and geometrical transformations used to model an antenna pointing at the sky.
\S\ref{sec::antennas} and \S\ref{sec::convolution} describe the methodology that was used to model the polarizing properties of an antenna and its coupling to the sky.
\S\ref{sec::pisco} describes the algorithm used by PISCO, as well as a prototype implementation. 
\S\ref{sec::validation} presents validation tests on the prototype implementation of PISCO. 
\S\ref{sec::realistic_cmb_experiment} provides an example application of PISCO to the Cosmology Large Angular Scale Surveyor, CLASS (\cite{2016SPIE.9914E..1KH}). 
The paper concludes with a brief discussion of the results obtained in \S\ref{sec::conclusions}.

\section{Coordinates}
\label{sec::coordinate-systems}

\begin{figure*}
	\centering
	\includegraphics[width=0.47\linewidth]{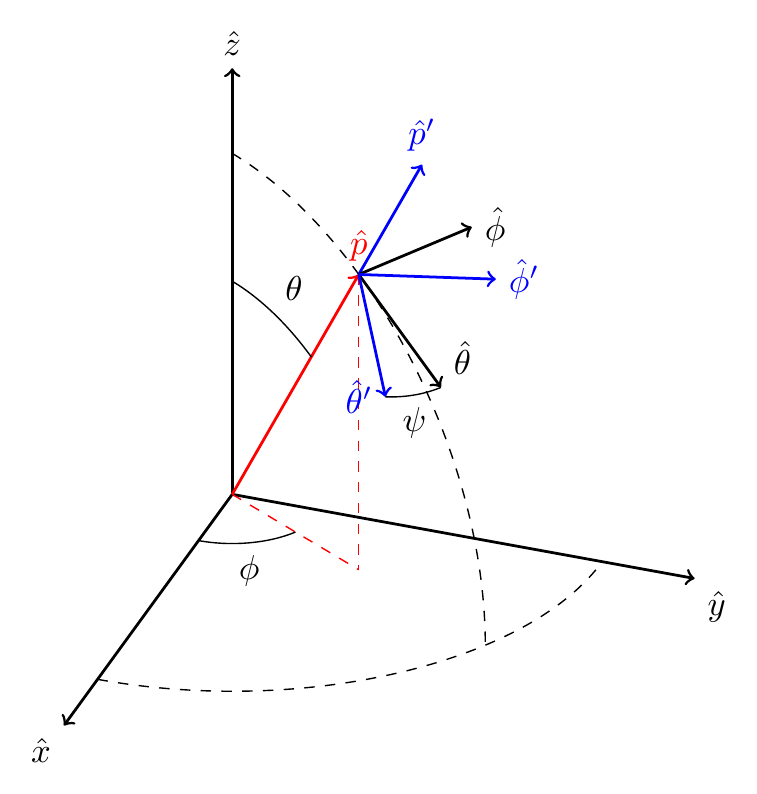}	
	\includegraphics[width=0.47\linewidth]{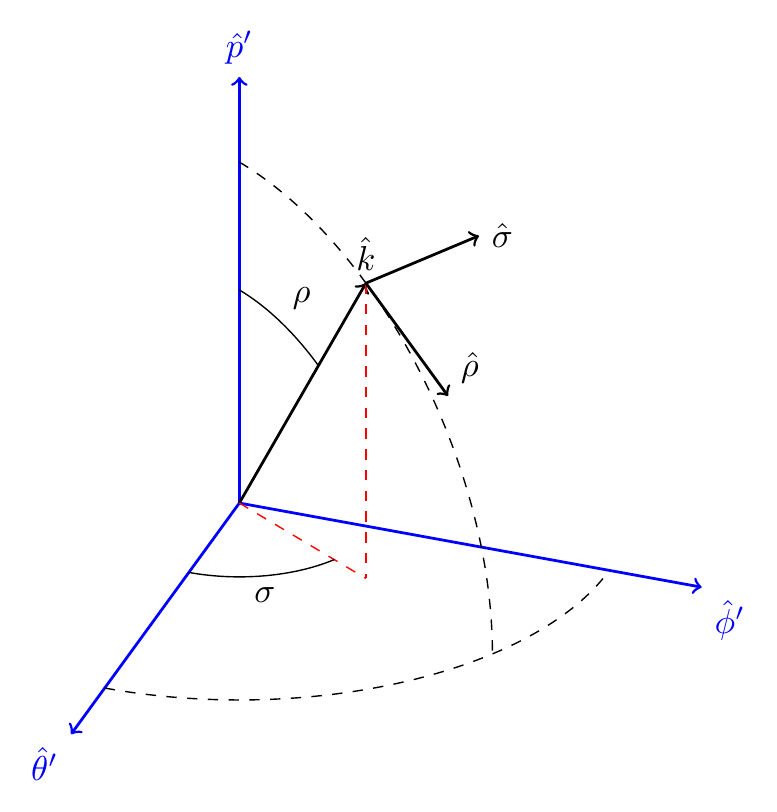}
	\caption{Left panel: sky basis. The sky basis is a generic spherical coordinate system. $\hat{x}$, $\hat{y}$ and $\hat{z}$ form an orthonormal basis. Unit vector $\hat{p}$ is defined by its spherical coordinates, co-latitude $\theta$ and longitude $\phi$. Co-latitude increases from the north pole towards the south pole. Longitude increases from west to east. Tangent vectors at $\hat{p}$, $\hat{\theta}$ and $\hat{\phi}$, can be rotated around $\hat{p}$ by angle $\psi$ to generate vectors $\hat{\theta}'$ and $\hat{\phi}'$. Note an observer looking towards the sky along $\hat{p}$ will measure angle $\psi$ as increasing clockwise from South. Right panel: the antenna basis is built using spherical unit vectors $\hat{\theta}'$, $\hat{\phi}'$ and $\hat{p}'$, analogous to Cartesian unit vectors $\hat{x}$, $\hat{y}$ and $\hat{z}$. The orientation of the antenna basis with respect to the sky basis is completely determined by the pointing vector $\hat{p}$ and rotation angle $\psi$. A unit vector $\hat{k}$ in the antenna basis is described by $\rho$ (co-latitude) and $\sigma$ (longitude). Tangent vectors at $\hat{k}$ are $\hat{\rho}$ and $\hat{\sigma}$. }
	\label{fig::sky_basis} 
\end{figure*}

Since PISCO performs the convolution of the polarized antenna response with the sky in the spatial domain, it makes extensive use of coordinate transformations. These operations can be described by the use of two complementary spherical coordinate systems, which are described in Figure \ref{fig::sky_basis}. The first coordinate system corresponds to the \textsl{sky basis}. Unit base vectors of these coordinate system are $\hat{x}$, $\hat{y}$ and $\hat{z}$. For convenience, the antenna pointing in the sky basis will be described as a 3-tuple $\bar{q}$, so that an antenna aiming at co-latitude $\theta_0$ and longitude $\phi_0$, with position angle $\psi_0$ has a pointing 

\begin{equation}
\begin{aligned}
\bar{q}_0= (\theta_0,\phi_0,\psi_0) \, \rm{.}
\end{aligned}
\end{equation}

Then the \textsl{pointing direction}, denoted by vector $\hat{p}_0$, can be expressed as a linear combination of base vectors and spherical coordinates $(\theta_0,\phi_0)$ via

\begin{equation}
\begin{aligned}
\hat{p}_0 = \sin(\theta_0)\cos(\phi_0) \hat{x} + \sin(\theta_0)\sin(\phi_0) \hat{y} + \cos(\theta_0) \hat{z} \, \rm{.}
\end{aligned}
\label{eq::p_sky_basis}
\end{equation}

The vectors $\hat{\theta}_0$ and $\hat{\phi}_0$ are computed using

\begin{eqnarray}
\begin{aligned}
\hat{\theta}_0 &=&  \cos(\theta_0) \cos(\phi_0) \hat{x} + \cos(\theta_0)\sin(\phi_0) \hat{y} - \sin(\theta_0) \hat{z} \\
\hat{\phi}_0   &=&              -\sin(\phi_0) \hat{x} +             \cos(\phi_0) \hat{y} \, \rm{.}
\end{aligned}
\label{eq::tangent_sky_basis}
\end{eqnarray}

These vectors can be used to build a second coordinate system, the \textsl{antenna basis}. Given an antenna pointing $\bar{q}_0$, the antenna basis base vectors can be written in terms of sky basis coordinates as 

\begin{eqnarray}
\hat{p}'_0      &=&  \hat{p}_0 \\
\hat{\theta}'_0 &=&  \cos(\psi_0)\hat{\theta}_0 + \sin(\psi_0)\hat{\phi}_0 \\
\hat{\phi}'_0   &=& -\sin(\psi_0)\hat{\theta}_0 + \cos(\psi_0)\hat{\phi}_0 \, \rm{.}
\label{eq::antenna_base_vectors}
\end{eqnarray}

In the antenna basis, coordinates analog to sky basis co-latitude and longitude are $\rho$ and $\sigma$, respectively. As in equation \ref{eq::p_sky_basis}, a vector $\hat{k}$ can be similarly written in terms of antenna basis coordinates as

\begin{equation}
\begin{aligned}
\hat{k}       &=&  \sin(\rho)\cos(\sigma)\hat{\theta}'_0 + \sin(\rho)\sin(\sigma) \hat{\phi}'_0 + \cos(\rho) \hat{p}'_0 
\end{aligned}
\end{equation}

\noindent
while vectors analog to the ones described by equation \ref{eq::tangent_sky_basis} are

\begin{eqnarray}
\begin{aligned}
\hat{\rho}    &=&  \cos(\rho)\cos(\sigma)\hat{\theta}'_0 + \sin(\rho)\sin(\sigma) \hat{\phi}'_0 - \sin(\rho) \hat{p}'_0 \\
\hat{\sigma}  &=& -\sin(\sigma)\hat{\theta}'_0 + \cos(\sigma)\hat{\phi}'_0 \, \rm{.}
\end{aligned}
\end{eqnarray}

\begin{figure}
	\centering
	\includegraphics[width=1.0\linewidth]{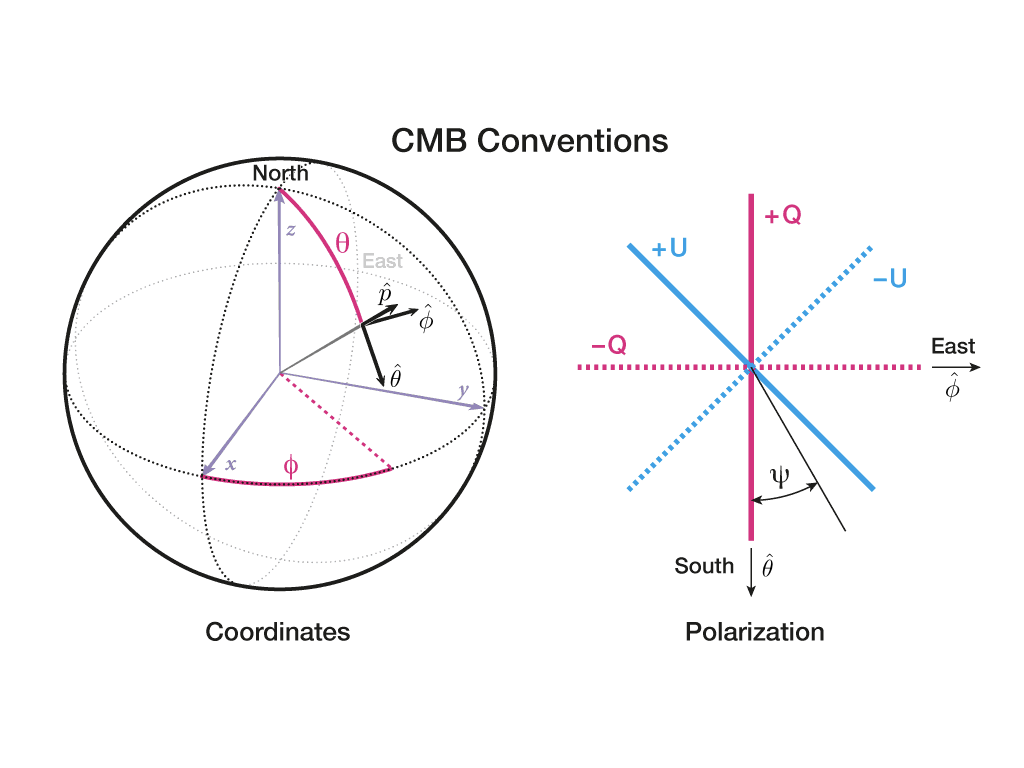}
	\caption{Figure showing the sign conventions of Stokes parameters $Q$ and $U$ used by the CMB community. In this figure, $\psi$ is the angle between the antenna basis base unit vector $\hat{\theta}'$ and $\hat{\theta}$. Positive $Q$ if the polarization vector is aligned with $\hat{\theta}$ (North-South direction), negative $Q$ if the polarization vector is aligned with $\hat{\phi}$ (East-West direction), positive $U$ is aligned with $(\pm(\hat{\phi} + \hat{\theta})/\sqrt{2} )$ (North/West-South/East direction), and negative $U$ is aligned with $(\pm(\hat{\phi} - \hat{\theta})/\sqrt{2} )$ (North/East-South/West direction). The right-most panel of this figure corresponds to an observer looking towards Earth.  This figure is adapted from one that was found on NASA's \href{https://lambda.gsfc.nasa.gov/product/about/pol_convention.cfm}{LAMBDA website} and was originally published in \cite{healpixDocs}}
	\label{fig::cmbcoordconvention}
\end{figure}

In many situations, antennas are equipped with polarization sensitive devices. The direction on the sky for which the device has maximum sensitivity to linearly polarized radiation is called co-polarization, denoted by $\hat{e}_{\co}$. The perpendicular direction, known as cross-polarization, is denoted by $\hat{e}_{\cx}$.
Because polarization is defined in the sky basis (see Figure \ref{fig::cmbcoordconvention}), a polarization sensitive antenna must compensate for the apparent rotation of its own polarization basis with respect to the sky. This can be accomplished by rotating the incoming Stokes vector by the angle between $\hat{e}_{\co}$ and $\hat{\theta}$, namely

\begin{equation}
\chi(\rho,\sigma) = \arctan \left( \frac{ |\hat{e}_{\co} \times \hat{\theta}| }{ \hat{e}_{\co} \cdot \hat{\theta} } \right) \, \rm{.}
\label{eq::psi}
\end{equation}

See Appendix A for details on the definition of co-polarization and cross-polarization that PISCO uses and the computation of $\chi(\rho,\sigma)$ using spherical trigonometry.

\section{A polarized model for antennas}
\label{sec::antennas}

The properties of an antenna equipped with detectors that are insensitive to polarization can be fully characterized by its beam. The beam is defined in terms of the antenna angular power density distribution $U(\rho,\sigma)$ via

\begin{equation}
\begin{aligned}
\mathit{b}(\rho, \sigma) = \frac{ U(\rho, \sigma) }{ \mathrm{max}\left[ U(\rho,\sigma) \right] }  =  \frac{ U(\rho, \sigma) }{ U(0,0) } \, .
\end{aligned}
\label{eq::beam_def}
\end{equation}

\noindent
From this definition of the beam, a standard quantification of an antenna's ability to direct power to a particular region of the sky is given by the beam solid angle $\Omega$, calculated as

\begin{equation}
\begin{aligned}
\Omega = \int_{4\pi} \mathit{b}(\rho,\sigma) \, \mathrm{d} \Omega \, .
\end{aligned}
\label{eq::omega_def}
\end{equation}

\noindent
These concepts fully characterize a lossless antenna in the case it is not sensitive to polarization, while modern CMB experiments, aim at measuring E-modes and B-modes. It then becomes necessary to introduce a more general treatment of the antenna, so as to include its polarizing properties. 

Many CMB experiments use polarization sensitive bolometers (PSBs) as detection devices, which are pairs of bolometers placed at the focus of the optical chain. By construction, each bolometer is more sensitive to a particular orientation of incoming electric fields. Given a PSB consisting of a pair of detectors $(a,b)$ whose polarization sensitive axes are aligned at angles $\zeta$ and $\zeta - 90^{\circ}$ with respect to the antenna basis unit vector $\hat{\theta}'_0$, we can define beam center polarization unit vectors for each detector as

\begin{align}
\hat{e}_{a,{\co}}    &= \quad \cos(\zeta)\hat{\theta}'_0 + \sin(\zeta)\hat{\phi}'_0 \\
\hat{e}_{a,{\cx}}    &= \quad \sin(\zeta)\hat{\theta}'_0 - \cos(\zeta)\hat{\phi}'_0 \\
\hat{e}_{b,{\co}}    &= \quad \sin(\zeta)\hat{\theta}'_0 - \cos(\zeta)\hat{\phi}'_0 \\
\hat{e}_{b,{\cx}}    &= -\cos(\zeta)\hat{\theta}'_0 - \sin(\zeta)\hat{\phi}'_0 \, \rm{.} 
\label{eq::copol_bolo}
\end{align}

The polarizing properties of an antenna can be obtained by combining the concept of a beam with the Mueller matrix formalism. Mueller matrices are widely used to quantify the effect that an optical element has on the polarization state of incoming light. This process is modeled by the multiplication of $4\times4$ matrix $\mathbf{M}$ with a Stokes vector $S_{\mathrm{in}}$, such that the Stokes vector of radation that has interacted with the optical element becomes

\begin{equation}
\begin{aligned}
S_{\mathrm{out}} = \mathbf{M} S_{\mathrm{in}} \, .
\end{aligned}
\end{equation}

\noindent
As described in the work of \cite{piepmeier_long_njoku_2008} and \cite{2007MNRAS.376.1767O}, antennas can be modeled using Mueller matrices. PISCO uses the formalism presented in \cite{2007MNRAS.376.1767O}, since it is more suitable to be applied to CMB experiments. We note the original paper names ``beam Mueller fields'' to this extended definition of a antenna beam. To emphasize the multi-dimensional nature of this mathematical entity, in this work we refer to it as a beam tensor, or \textsl{beamsor} for short.

A beamsor can be interpreted as a field of Mueller matrices such that for each direction $(\rho,\sigma)$, there is an associated Mueller matrix that quantifies the coupling between the antenna and a Stokes vector coming from $(\rho,\sigma)$. We will denote a beamsor by letter $\mathbf{B} = \mathbf{B}(\rho,\sigma)$. At every antenna basis direction, $\mathbf{B}$ is a $4\times4$ matrix in the form

\begin{equation}
\begin{aligned}
\mathbf{B}(\rho,\sigma) = \frac{1}{\tilde{\Omega}}
\begin{bmatrix}
B_{TT} & B_{QT} & B_{UT} & B_{VT}\\
B_{TQ} & B_{QQ} & B_{UQ} & B_{VQ}\\
B_{TU} & B_{QU} & B_{UU} & B_{VU}\\
B_{TV} & B_{QV} & B_{UV} & B_{VV}
\end{bmatrix}
\end{aligned}
\label{eq::beamsor}
\end{equation}

\noindent
where $\tilde{\Omega}$ is a normalization factor given by

\begin{equation}
\begin{aligned}
\tilde{\Omega} = \int_{4\pi} B_{TT}(\rho,\sigma) \, \mathrm{d} \Omega
\end{aligned}
\end{equation}

\noindent
and the elements of $\mathbf{B}$ are defined in the work of \cite{2007MNRAS.376.1767O}. Note that in the case of no cross polarization and perfectly matched beams, the beamsors reduce to the Kronecker product 

\begin{equation}
\mathbf{B}(\rho,\sigma) = \mathit{b}(\rho,\sigma) \otimes \mathbf{I}{}
\end{equation}

\noindent
where $\mathbf{I}$ is an identity matrix. 

\section{Measuring the sky with polarization sensitive antennas}
\label{sec::convolution}
	
\subsection{Continuous case}

In order to model the process by which a polarization sensitive detector transforms electromagnetic radiation into current or voltage, we used the formalism described in \cite{2007A&A...470..771J}. In Mueller matrix space, a partially polarized, total power detection device corresponds to the following row-vector 

\begin{equation}
\begin{aligned}
\tensor{D}{}(\zeta,\epsilon,s) = \frac{s}{2} \left[(1 + \epsilon), (1 - \epsilon)\cos(2\zeta), (1 - \epsilon)\sin(2\zeta), 0 \right]
\end{aligned}
\label{eq::M_pol}
\end{equation}

\noindent
where $1 - \epsilon$ is the polarization efficiency, $s$ is the voltage responsivity of the detector and $\zeta$ was defined in \S\ref{sec::antennas}. The process of taking a total power measurement on a Stokes vector $\tensor{S}{^{i}}$ can then be modeled as a dot product which, using Einstein's summation convention, yields

\begin{equation}
\begin{aligned}
d = \tensor{D}{_{i}} \tensor{S}{^{i}} \, .
\end{aligned}
\label{eq::total_power_measurement}
\end{equation}

Stokes vector $\tensor{S}{^{i}}$ is, in turn, the convolution between the antenna beamsor and the polarized sky. The complete expression describing the measurement taken by a linearly polarized, total power detector coupled to an antenna pointing along $\hat{p}$ with position angle $\psi_0$ becomes

\begin{equation}
\begin{aligned}
d(\bar{q}_0) =  \int_{4\pi} \tensor{D}{_{i}}(\chi,\epsilon,s) \tensor{\Lambda}{^{i}_{l}}(\chi)  \tensor{B}{^{l}_{j} } \left[ \tensor{\Lambda}{^{j}_{m}}(-\chi) \tensor{S}{^{m}} \right ] \, \mathbf{d} \Omega 
\end{aligned}
\label{eq::pisco_equation_cont}
\end{equation}

\noindent
where $(i,j) = T,Q,U,V$ and all beamsor elements $\tensor{B}{^{l}_{j} }$ are aligned with $(\hat{p},\psi_0)$. Rotation of the detector polarization basis to the sky basis is carried out using $\chi$ in the argument to $D$, and by $\mathbf{\Lambda}$, which is a matrix field in the form

\begin{equation}
\begin{aligned}
\mathbf{\Lambda}(\chi) =
\begin{bmatrix}
1  & 0 & 0 & 0\\
0  & \cos(2\chi) & -\sin(2\chi) & 0\\
0  &\sin(2\chi) & \cos(2\chi) & 0\\
0  & 0 & 0 & 1
\end{bmatrix}  \, \rm{.}
\end{aligned}
\label{eq::lambda_operator}
\end{equation}

\subsection{Pixelated case}
\label{sec::pixel_conv}

In order to calculate the result of equation \ref{eq::pisco_equation_cont} using a computer, we can no longer use continuous distributions, so we need to use pixelated versions of both the beamsor and sky model. Both the sky model $S$ and beamsor $\mathbf{B}$ must then be transformed into a matrix of $N_b$ (number of pixels in the beamsor) and $N_s$ (number of pixels in the sky model) entries, respectively. We will denote the $k$-th pixel of the pixelated beamsor $\mathbf{B}$ (sky $S$) as $\tensor[_k]{B}{}$ ($\tensor[_k]{S}{}$). This way, the element $(i,j)$ of beamsor $\mathbf{B}$ at pixel $k$ is denoted $\tensor[_k]{B}{^i_j}$. With the above in mind, we can write equation \ref{eq::pisco_equation_cont} for the pixelated case as

\begin{equation}
\begin{aligned}
d(\bar{q}_0) = \
\sum_{k=1}^{N_s} \
\tensor[_k]{D}{_{i}}(\chi,\epsilon,s) \
\tensor[_k]{\Lambda}{^{i}_{l}}(\chi) \
\tensor[_k]{B}{^l_j} \
\tensor[_k]{\Lambda}{^{j}_{m}}(-\chi) \
\tensor[_k]{S}{^m}  \, \rm{.}
\end{aligned}
\label{eq::discrete_beam_conv_tensor}
\end{equation}

\noindent
where $\mathbf{B}$ has been properly re-pixelated via interpolation, so as to keep track of the actual position of the beam on the sky according to $\bar{q}_0$ using the equations presented in Appendix A. It is worth noting that this expression operates on a pixelated sky, so power from angular scales smaller than a sky pixel is greatly suppressed by the corresponding pixel window function.

\section{PISCO}
\label{sec::pisco}

The \textbf{PI}xel \textbf{S}pace \textbf{CO}nvolver (PISCO) is a tool with the capability of generating synthetic Time Ordered Data (TOD) provided a model for the beamsor, the scanning strategy of the mission and a sky model. In this section, we present a pathfinder implementation of PISCO that exploits the massively parallel architecture of modern GPU systems.

\subsection{General description}

\begin{figure}
	\centering
	\includegraphics[width=0.6\linewidth]{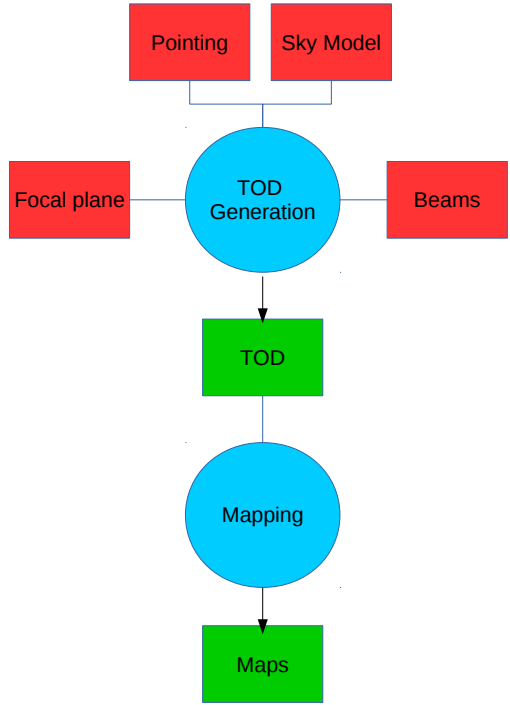}
	\caption{Basic flow of a typical PISCO simulation pipeline. Red polygons show the required user input. PISCO uses this input and produces TOD (green polygon). This TOD stream is calculated using equation \ref{eq::discrete_beam_conv_tensor} for all pointing directions. TOD can then by sent into a mapper and, finally, to a power spectra estimator tool. PISCO does not compute pointing nor produces maps from TOD by itself; these tasks are left to external programs.}
	\label{fig::pisco_flow}
\end{figure}

PISCO is the software tool in charge of generating mock TOD given a beamsor, a sky model and a scanning strategy. A diagram showing the general workings of PISCO is shown in Figure \ref{fig::pisco_flow}. PISCO receives as input a sky model in the form of 4 maps representing Stokes parameters $I,Q,U$ and $V$, a beamsor, pointing and focal plane information. PISCO stores the beamsor elements and sky model as HEALPix (see \cite{2005ApJ...622..759G}) maps. HEALPix was chosen because it is widely used among the CMB community, and because it naturally handles the closed surface topology of the sphere. HEALPix also provides equal area pixels, which is a desirable feature when computing convolution in pixel space. The focal plane specifications are only needed if multiple detectors are being included in the pointing stream, as PISCO needs the angle $\zeta$ of each detector to compute equation \ref{eq::discrete_beam_conv_tensor}. All the inputs are sent to the TOD generation function, which returns the data streams. At this point, the data can either be saved to disk or sent to a map-making code. This last step is preferred as, usually, input-output operations are time consuming. Finally, maps can be analyzed using external tools to calculate the power spectra.

\subsection{Implementation using CUDA}

\begin{figure}
	\begin{center}
		\includegraphics[width=0.8\linewidth]{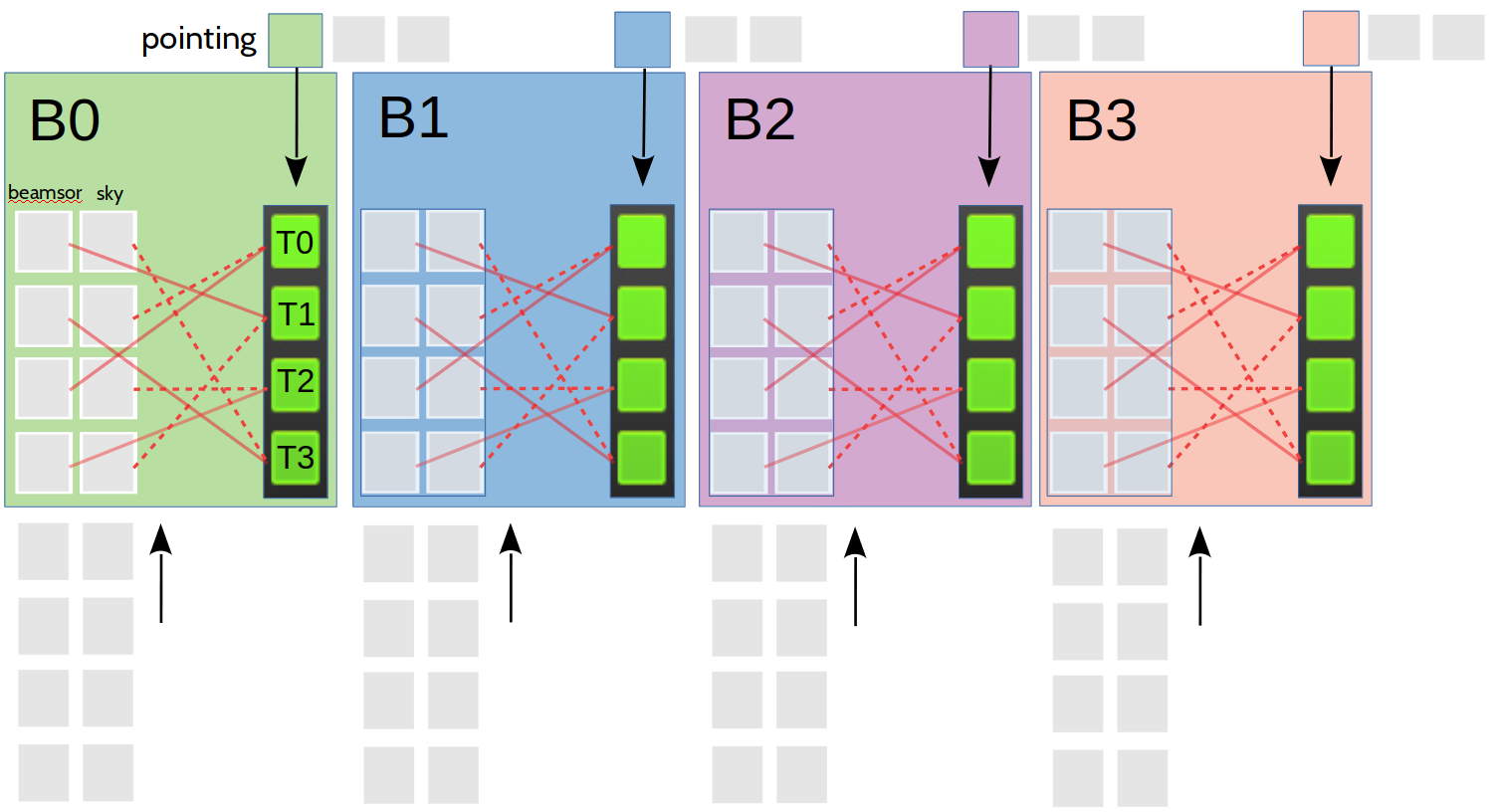}
		\caption{Parallelization scheme. This figure shows the case of PISCO executing in 4 blocks ($B=4$), with four threads per block ($T=4$). Arrays with beamsor and sky elements are at the bottom. Each block has access to four beamsor and sky pixels (gray boxes inside colored boxes) and one pointing entry (colored small boxes) at a time. Green boxes represent the multiplication process of a single beamsor pixel with a single sky pixel. This includes rotating the sky pixel to the detector polarization basis, and computing the re-pixelization of $B$ at the corresponding coordinates. Solid and dotted red lines represent the complex memory access pattern generated by this process. Every thread within a block writes its result to shared memory space. When a thread finishes its computation, it waits until all threads have finished and a reduction on the shared memory space is performed across all blocks. This process is repeated for every pointing. At the end of the procedure, each block has computed the convolution of a beamsor with the sky for a particular pointing, and every shared memory space of the block has the corresponding result. These results are collected into the GPU global memory, which is then transferred back to CPU (host) memory.}
		\label{fig::pisco_diagram}
	\end{center}
\end{figure}

GPUs allow for substantial acceleration of algorithms that perform a large amount of independent operations. TOD generation using equation \ref{eq::discrete_beam_conv_tensor} presents an optimal application case because all operations are independent of each other. In this work, we used the Compute Unified Device Architecture framework from NVIDIA to implement the TOD generation routine. The reader is referred to \cite{sanders2010cuda} for an excellent description of CUDA and associated capabilities.

To better understand how the parallelism in \ref{eq::discrete_beam_conv_tensor} can exploited, consider the process of synthesizing $N_T$ measurements using a CUDA grid of $B$ blocks and $T$ threads. Consider each measurement to have an associated pointing $\bar{q}_t$ with $t=0..N_T$. PISCO performs a double parallelization scheme: the ``slow'' loop ($L1$) scans the pointing stream and associates every block to a pointing $\bar{q}_t$. A second, ``faster'' loop ($L2$), iterates over a list of pixels, which correspond to sky pixels that are ``inside'' the beamsor extension. This list of pixels is constructed in advance and then transferred to the GPU. $L2$ executes $T$ operations in parallel. Great care was taken to ensure no race conditions arise when multiple threads try to read (write) from (to) the same memory address. As every block executes $T$ convolution operations in parallel, and the CUDA grid runs $B$ simultaneous blocks, the parallelism is $B \times T$. Furthermore, if $G$ GPUs are available, the computation can be distributed among them, increases the parallelism to $G \times B \times T$. A graphical description of this process is shown in Figure \ref{fig::pisco_diagram}.

\subsection{Performance}

\subsubsection{Benchmark results}

To gain insight regarding the performance of the algorithm, we performed a simple benchmark where a single PSB scanned the whole sky with a Gaussian beam of FWHM $=1.5^\circ$. We restricted the benchmark to Gaussian beams only, but emulated the presence of sidelobes by increasing the number of sky pixels involved in the convolution on every run. The benchmark consisted of timing 2 detectors with the same beams scanning the complete sky at 3 different beam orientation angles ($0^\circ,45^\circ$ and $90^\circ$). The pointing of every detector was set to aim at the center of every sky pixel, effectively setting the number of samples in the TOD to 3 times the number of pixels on the sky, per detector. We used a polarized CMB map of \texttt{NSIDE} = 128 (196608 pixels) as input, which set the number of samples in the synthetic TOD to 1179648 ($2\times3\times196608$). The NSIDE parameter of the beam was set to 512. The test was run on a single GTX 1080 card and a node equipped with two Intel Xeon E5-2610 processors (10 physical cores and 20 threads per processor) and 256 GB of DDR4 RAM. Unfortunately, the current code architecture does not allow for a simple way of measuring performance in FLOPS, which is a challenging task on its own given the TOD generation routine performs both integer and floating point mathematical operations. For this reason, we report execution wall-time as a metric for performance.

Table \ref{tab::performance} shows the scaling of wall-time with the maximum radial extent of the beam (denoted as $\rho_\mathrm{max}$) was found to be roughly quadratic. Since the number of pixels in a disc depends on the pixelization scheme, we report the following relation between wall time and the number of sky pixels evaluated per convolution for this particular test.

\begin{equation}
\Delta t =  4.17 N_p - 575.5
\label{eq::pisco_walltime}
\end{equation}

\noindent
where $\Delta t$ is wall time in milliseconds and $N_p$ is the number of sky pixels. Auxiliary runs showed that $\Delta t$ scales linearly with the number of pointing directions involved.

The behavior of this scaling is expected from the architecture of the algorithm (see Figure \ref{fig::pisco_diagram}), and can be ameliorated by using a distributed computing architecture with multiple GPUs. Another free parameter, the \texttt{NSIDE} parameter of the beam, was found to have a minor effect on execution time. The degradation caused by using a finer beam is believed to be a consequence of the GPU having less free memory to store the evaluation grid, which reduces the number of convolutions it can calculate in parallel.

\begin{figure}
	\centering
	\includegraphics[width=0.75\textwidth]{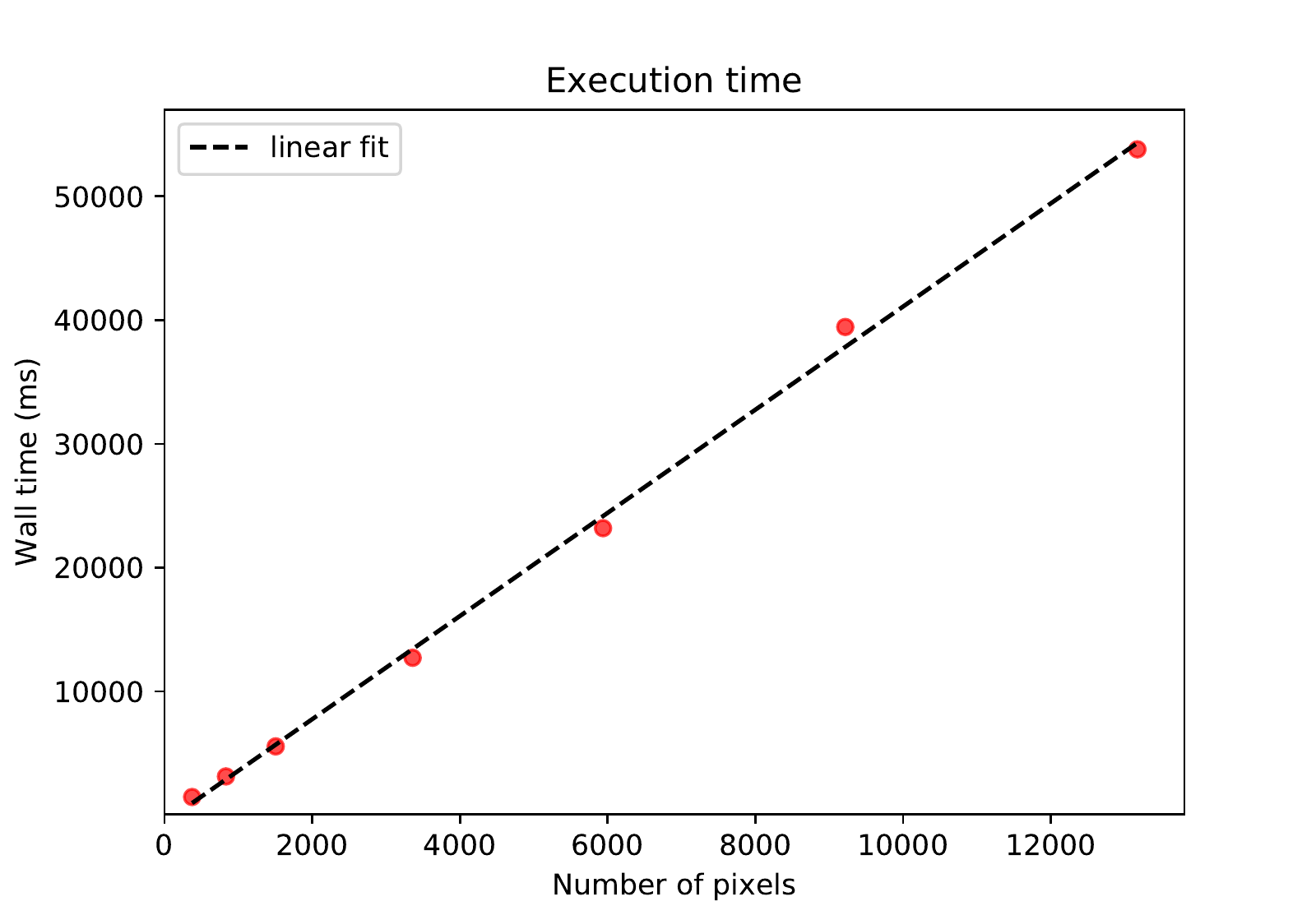}
	\caption{Plot showing data from table \ref{tab::performance}. The red circles correspond to the data, while the black, dashed line corresponds to the best linear fit. The scaling of wall-time with the number of pixels involved in the convolution is linear, as expected from the nature of the algorithm. }
	\label{fig::walltime}
\end{figure}

\begin{table}[]
	\centering
	\begin{tabular}{ccc}
		$\rho_{\rm{max}}$ (${}^\circ$) & Pixels & Wall time (ms) \\
		5.0 & 374 & 1471.41 \\
		7.5 & 832 & 3135.19 \\
		10.0 & 1506 & 5560.25 \\
		15.0 & 3360 & 12719.56 \\
		20.0 & 5938 & 23190.68 \\
		25.0 & 9218 & 39447.70 \\
		30.0 & 13174 & 53800.74
	\end{tabular}
	\caption{Table showing wall-times taken by a whole-sky convolution of a single PSB. $\rho_{\rm{max}}$ corresponds to the ``cut-off'' angle that is taken into account by the core convolution algorithm. The central column corresponds to the maximum number of sky pixels involved in the convolution. The third column corresponds to wall time, including the assembly of the beam interpolation grid, copying memory between the host and the GPU, as well as the wall-time of the main convolution routine. We note that, because of the limited amount of RAM in the GPU, runs with $\rho_{\rm{max}} > 20^\circ$ were performed in blocks in a purely serial manner.}
	\label{tab::performance}
\end{table}

\subsubsection{Discussion and future improvements}

Algorithms like beamconv (see \cite{2018arXiv180905034D}) and the smoothing routine present in the healpy package, work in harmonic space. This means their execution times are much faster as they rely on Discrete Fast Fourier Transform techniques. It is worth noting that, in theory, the nature of pixel space permits the inclusion of other features in the simulation (beam ghosting, sharp sidelobes or
time-dependent phenomena) at a relatively small additional computational cost. We believe this makes PISCO a good complement to other software tools that perform similar tasks.

The current implementation must calculate the list of sky pixels involved in each convolution, for all pointing directions, before the CUDA routine is launched. While the wall-time associated with this operation is modest, the result must be kept in memory and transferred to the GPU, so that the associated buffer quickly becomes too large to be held in the VRAM. Currently, PISCO handles this situation by performing the generation of TOD in blocks to avoid memory overflow. In the test machine, computing and transferring the lists of pixels can take up to $13\%$ of the \textsl{overall} simulation wall-time. A solution to this problem has already been devised and will be implemented in future releases. Another drawback of the current implementation is the use of global memory to hold the beam tensor elements. Future releases will exploit data locality by making use of the CUDA texture memory pipeline (see \cite{sanders2010cuda} and \cite{2019AW...Wilt}). Finally, while the current implementation of PISCO was designed to execute in multiple GPU nodes, significant coding effort is required to provide the user with an easy to use interface. Experiments were performed emulating a multi-node system by making PISCO use all 3 GPUs of the machine. These tests showed an almost linear increase in performance, but more work is required in order to find the knee of the curve between performance and available GPUs.

\section{Code validation}
\label{sec::validation}

To validate the correctness of PISCO, we performed two sets of tests: a mock observation of a (polarized) point source and a simulation of an ideal CMB observation. In the case of the polarized point source observation, the output of PISCO was compared against the \texttt{healpy.smoothing} routine, a Python wrapper around HEALPix routines, which calculates the convolution of a polarized sky with a circularly symmetric Gaussian beam in $a_{\ell m}$ space. For the purposes of this test, we will consider the output of \texttt{smoothing} to be exact. In the case of the ideal CMB observation, the output from PISCO, after being compensated by the analytic beam window function, was compared against the power spectra of the input maps.

\subsection{Point source observations}

The simulated observation of a polarized point source was accomplished by the following steps

\begin{itemize}
    \item Build a beamsor with off-diagonal elements set to zero. Every $\tensor{B}{^i_i}$ element corresponds to a circular Gaussian beam with FWHM of $1.5^\circ$. 
    \item For simplicitly, we used a PSB with $s=1$ and $\epsilon=0$ for both detectors.
	\item Build a mock sky, with a single non-zero pixel at coordinates $(\theta_k,\phi_k)$. Three cases were run: unpolarized, pure Q polarization and pure U polarization..
	\item Set up a raster scan around $(\theta_k,\phi_k)$ for a detector with $\zeta=0$. Note that, in order to have full polarization coverage, the raster scan is repeated 3 times with angles $\psi_0 = 0^{\circ},45^{\circ},90^{\circ}$.
	\item Make maps of TOD generated by PISCO.
	\item Compare the result of applying \texttt{smoothing} to the single pixel map using the same beam model.
\end{itemize}

\subsubsection{Results}

\begin{figure}
	\centering
	\includegraphics[width=1.0\linewidth]{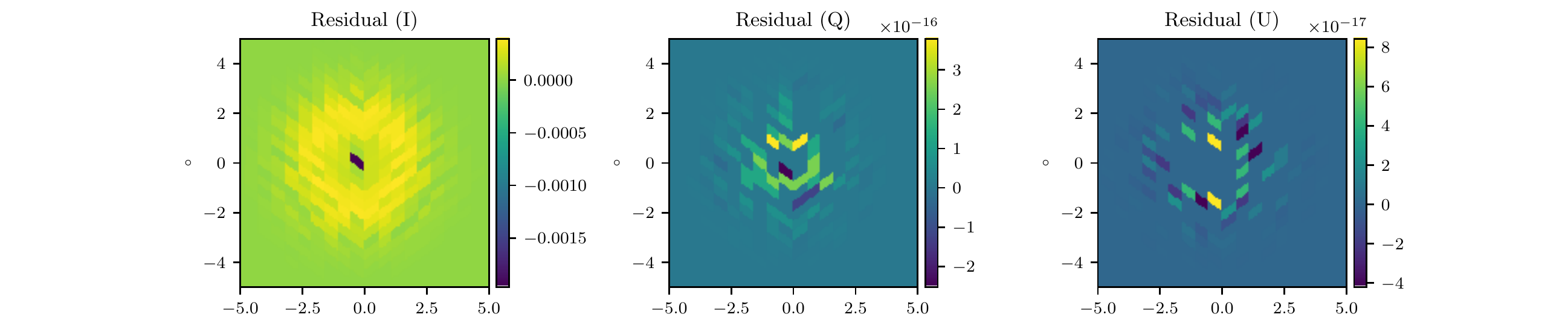}
	\includegraphics[width=1.0\linewidth]{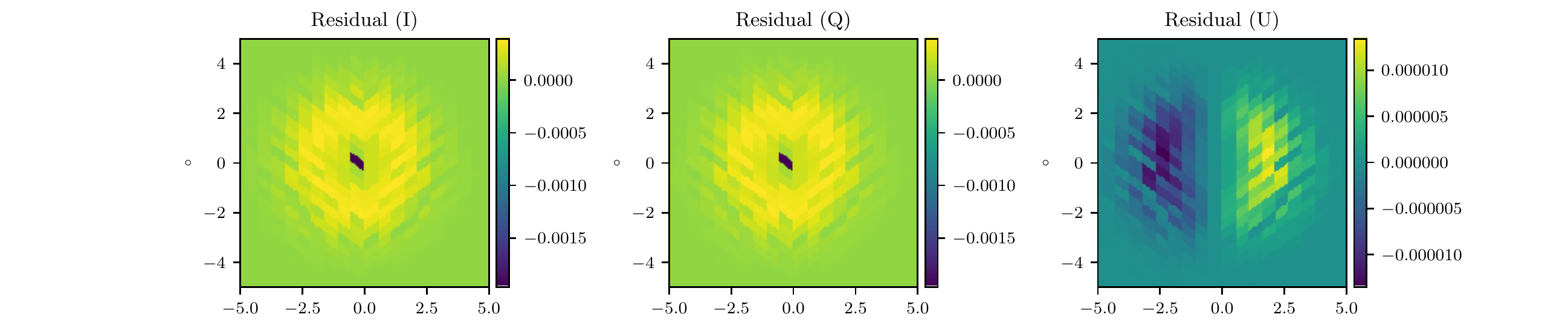}
	\includegraphics[width=1.0\linewidth]{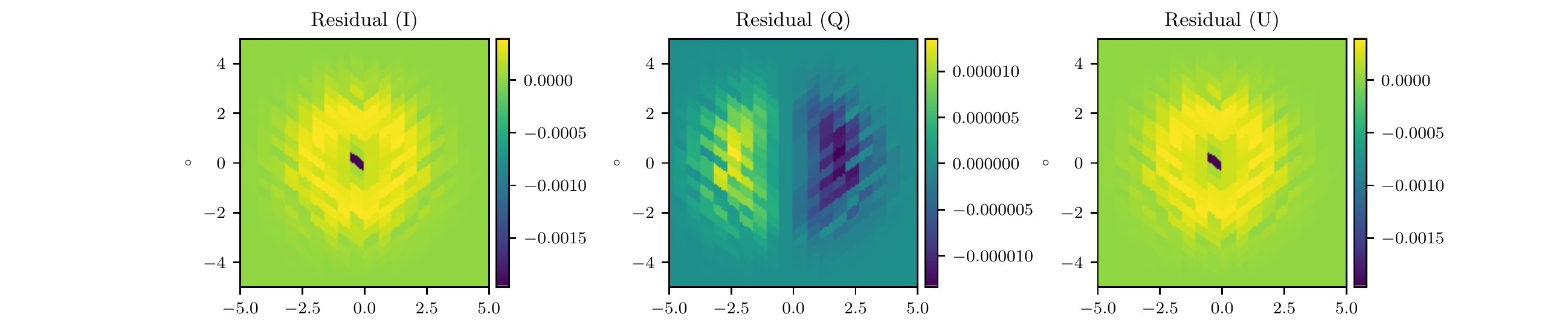}
	\caption{Plots showing the result of differencing a map with a point source convolved with a Gaussian beam using \texttt{smoothing} and the map generated from PISCO TOD. The input map used HEALPix pixelization with $\mathrm{\texttt{NSIDE}} = 256$, while the beamsor resolution was set to $\mathrm{\texttt{NSIDE}} = 1024$. The Gaussian profile used for the beamsor had a FWHM of $1.5^\circ$. All point sources were located at $45^\circ$ declination. The color scale is normalized to $1$, which is also the amplitude of the source. From top to bottom: residuals for the case of a point source with Stokes vector $S = (1,0,0,0)$, $S=(1,1,0,0)$ and $S=(1,0,1,0)$, respectively.}
	\label{fig::stokesqsource256beam1024dec45}
\end{figure}

Results for this validation test are shown in Figure \ref{fig::stokesqsource256beam1024dec45}. The results show that the flux is preserved to better than $0.2\%$. For the \texttt{NSIDE} 256 sky used in this test, doubling \texttt{NSIDE} of the beam from 1024 to 2048 does not improve the preservation of the flux significantly (from $0.2$\% to $0.04$\%) at the cost of quadrupling the amount of memory required to store the beamsor, while using a lower \texttt{NSIDE} for the beam degrades this considerably to $0.8$\%. We conclude that a 4:1 beam to sky \texttt{NSIDE} ratio is sufficient. We also checked for systematic effects driven by the finite machine precision of the computations, and found that leakage from temperature to polarization was consistent with zero to machine precision. The maximum residual in the P to P leakage is on the order of $0.1\%$. PISCO is also able to correctly account for intra-beam variations of the position angle $\chi$.

\subsection{Ideal CMB experiment}
\label{subsec::ideal_full_sky}

\subsubsection{Description}

The simulation of an ideal CMB experiment was accomplished by the following:

\begin{itemize}
    \item Build a beamsor with off-diagonal elements set to zero. Every $\tensor{B}{^i_i}$ element corresponds to a circular Gaussian beam with FWHM of $1.5^\circ$. 
    \item For simplicitly, we used a PSB with only one active detector for which $s=1$ and $\epsilon=0$.
    \item Build mock CMB whole sky maps with a tensor-to-scalar ratio $r=0.0$.
	\item Set up a scanning strategy to visit each pixel center at 3 different position angles $\psi_0 = 0^{\circ},45^{\circ},90^{\circ}$. 
	\item Make maps of TOD generated by PISCO. 
	\item Compare spectra generated from the maps with spectra of the input maps.
\end{itemize}

\subsubsection{Sky model}
\label{subsec::sky_model}

The input sky maps were generated using a combination of CAMB (see \cite{Lewis:2002ah}) to generate $C_\ell$ and \texttt{synfast} to generate maps from the $C_\ell$. The cosmological parameters used by CAMB are consistent with those reported by the \textsl{Planck} satellite collaboration (see \cite{2016A&A...594A..13P}). This procedure returns 3 CMB anisotropy maps, one for each Stokes parameter. The CMB is not expected to have Stokes $V$ polarization, so this field was explicitly set to zero. CAMB was configured to return a CMB with no primordial B-modes ($r=0$) and no lensing, as this last effect is expected to transform E-modes to B-modes. The resulting B-mode power spectrum is effectively zero at all angular scales. No foreground or other sources were added on top of the simulated CMB. All maps used for this simulation use the \texttt{HEALPix} pixelization scheme and were generated at a resolution parameter \texttt{NSIDE} $=512$.

\subsubsection{Scanning strategy and beam}

The pointing stream was built to convolve the NSIDE = 512 sky maps onto output maps of NSIDE = 128. We used as pointing every pixel position of a HEALPix map with NSIDE = 128, effectively sampling 
the input sky on 1/16th of its pixels. The goal of this mismatch between pointing and input sky resolution was to provide a more realistic simulation as, in reality, the input sky is not pixelated 
at all but it is the output maps that must be computed on a pixelated sphere. Every pixel was observed at its center, which is an important requirement that ensures the intra-pixel coverage does not affect the estimation of the power spectra at high $\ell$ (see Appendix B in \cite{2005poutanen}). Since only three hits per pixel at different values of $\chi$ are required to recover the polarization field of the CMB, the scanning was generated for a single detector with a polarization sensitive angle $\zeta=0$. Finally, the beam was set to be a circularly symmetric Gaussian with FWHM of $1.5^\circ$ and pixelated as a HEALPix map of \texttt{NSIDE} $=2048$.

\subsubsection{Power spectra}

Power spectra were calculated using \texttt{anafast}. No further post-processing of the power spectra was needed given that this simulated observation covers the whole sky, and hence no masking effects arise. The power spectra corresponding to maps that were generated using PISCO TOD were corrected by dividing by the equivalent beam transfer function of a circular Gaussian beam of FWHM $1.5^\circ$. Note that tools like \texttt{anafast} return $C_\ell$, whereas the literature usually shows $D_\ell$. In this work, we follow that convention:

\begin{equation}
D_\ell = \frac{\ell (\ell+ 1) }{2 \pi} C_\ell \, \rm{.}
\end{equation}

\subsubsection{Results}

\begin{figure}
	\centering
	\includegraphics[width=1\textwidth, trim = {1.2cm 0.0cm 1.2cm 0.0cm}, clip ]{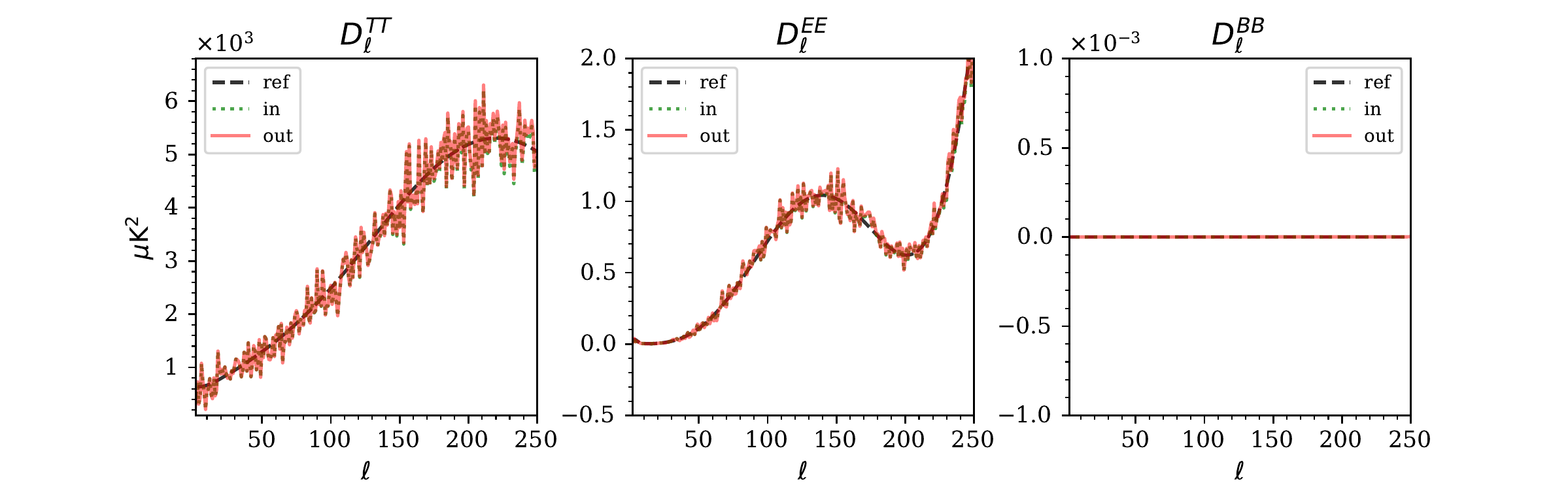}
	\includegraphics[width=1\textwidth, trim = {1.2cm 0.0cm 1.2cm 0.0cm}, clip ]{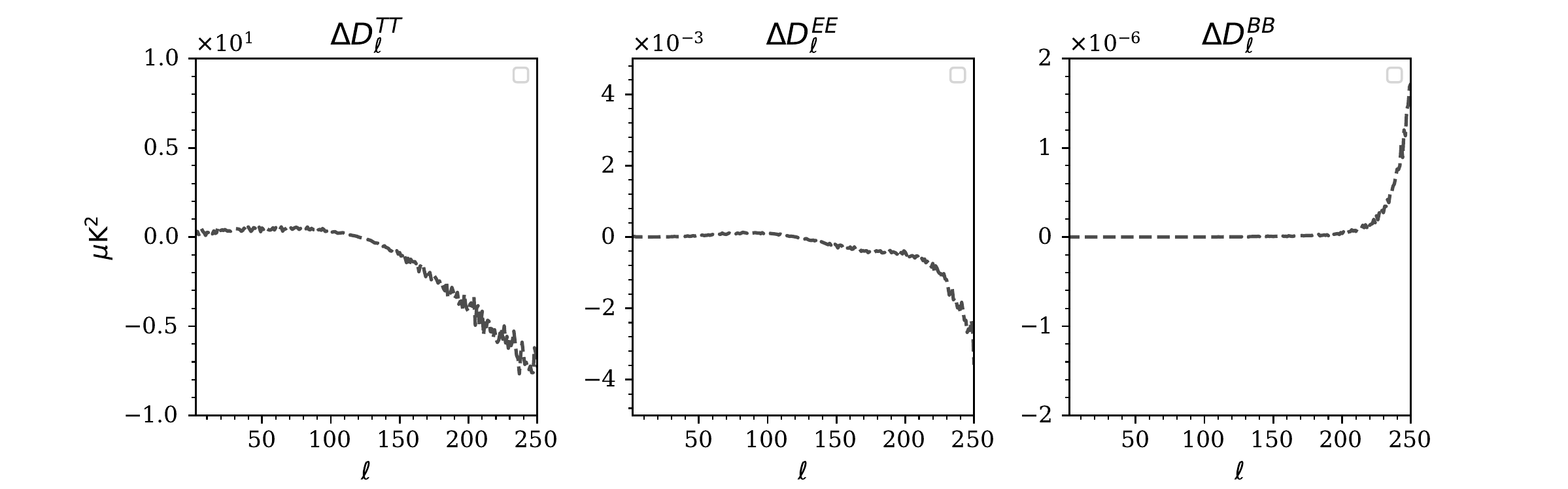}
	\caption{Upper plot: power spectra from PISCO generated TOD (``out'' label) for the whole sky CMB test with pointing on pixel centers. $D_\ell$ used to generate the input maps are the black, dashed curve (``ref'' label). The input sky corresponds to a CMB with $r=0.0$, and its power spectra are plotted in dotted green (``in'' label). The scale of the y-axis is $\mu\rm{K}^2$ for all three plots.  Visual inspection shows excellent agreement between PISCO and the input spectra, with no apparent leakage from E-modes to B-modes or from temperature to polarization. Bottom: difference between power spectra of PISCO generated TOD and the power spectra of the input map. Residuals in $TT$, $EE$ and $BB$ reach no more than $8\, \mu\mathrm{K}^2$, $0.004\, \mu\mathrm{K}^2$ and $2\, \times 10^{-6} \mu\mathrm{K}^2$, respectively. }
	\label{fig::pisco4wholesky}
\end{figure}

The main result of this simulation is shown in Figure \ref{fig::pisco4wholesky}. A single simulation took less than 25 seconds on the testing machine. We report no significant leakage from temperature to polarization, nor from E-modes to B-modes. It is worth noting that this process was repeated using several realizations of roughly the same cosmology, but with varying values of $r$. We found no evidence of leakage in any of these cases. 

\subsection{Ideal CMB experiment with more realistic pointing}
\label{subsec::ideal_full_sky_offsets}

The simulation described above used pixel centers as the telescope pointing. In an effort to provide more realism to the simulation, we dropped this constraint and performed a similar simulation to the one described in \S\ref{subsec::ideal_full_sky} by adding random offsets to the pointing with respect pixel centers, both in $\theta$ and $\phi$ as

\begin{eqnarray}
\phi_{i, \, \mathrm{new}} &=& \phi_{i} + \frac{ U( f \Omega ) }{ \sin{\theta_i} } \\
\theta_{i, \, \mathrm{new}} &=& \theta_{i} +  U(f\Omega) \, \rm{,}
\end{eqnarray}

\noindent
where $U(a)$ returns a uniformly distributed random number between $-0.5$ and $0.5$, $f$ is scaling factor and $\Omega$ is the average pixel size of a HEALPix map which, for \texttt{NSIDE} 128, is $0.4581^\circ$. For the purpose of this test, $f$ was set to $0.1$. 

\begin{figure}
	\centering
	\includegraphics[width=1\textwidth, trim = {1.2cm 0.0cm 1.2cm 0.0cm}, clip ]{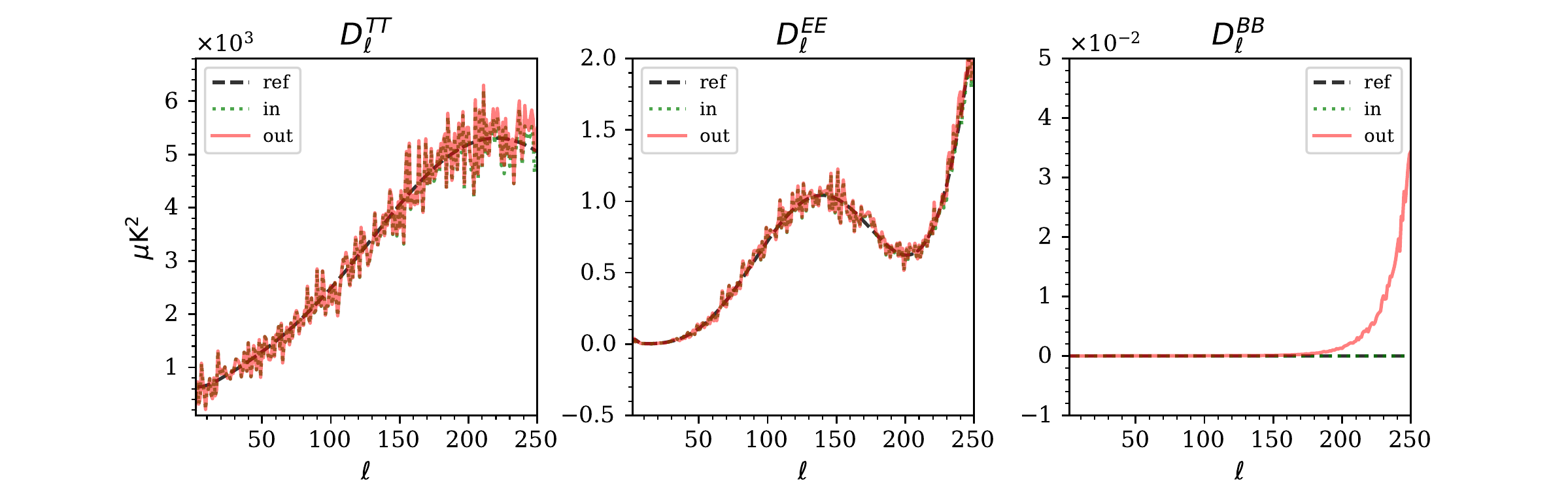}
	\includegraphics[width=1\textwidth, trim = {1.2cm 0.0cm 1.2cm 0.0cm}, clip ]{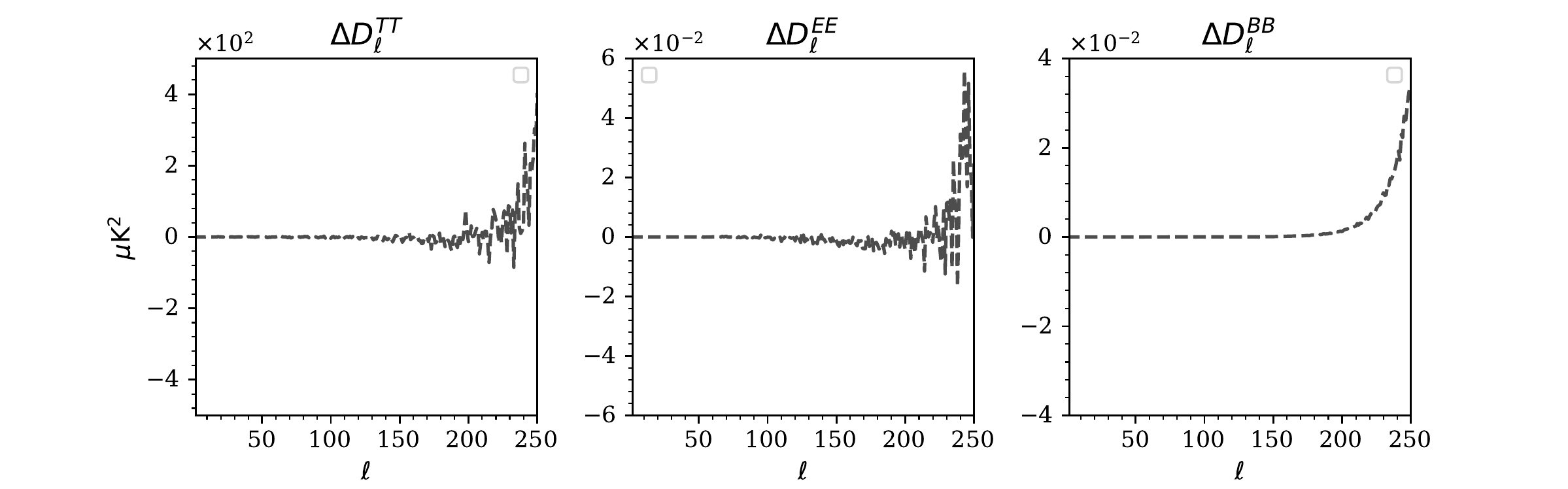}
	\caption{Upper plot: power spectra from PISCO generated TOD (``out'' label) for the whole sky CMB test with random pointing offsets from pixel centers. $D_\ell$ used to generate the input maps are the black, dashed curve (``ref'' label). The input sky corresponds to a CMB with $r=0.0$, and its power spectra are plotted in dotted green (``in'' label). The scale of the y-axis is $\mu\rm{K}^2$ for all three plots. Bottom: difference between power spectra of PISCO generated TOD and the power spectra of the input map. The spurious B-mode signal reaches $0.03 \, \mu\mathrm{K}^2$.}
	\label{fig::pisco4wholesky_random_offsets}
\end{figure}

\begin{figure}
	\centering
	\includegraphics[width=1\textwidth, trim = {1.2cm 0.0cm 1.2cm 0.0cm}, clip ]{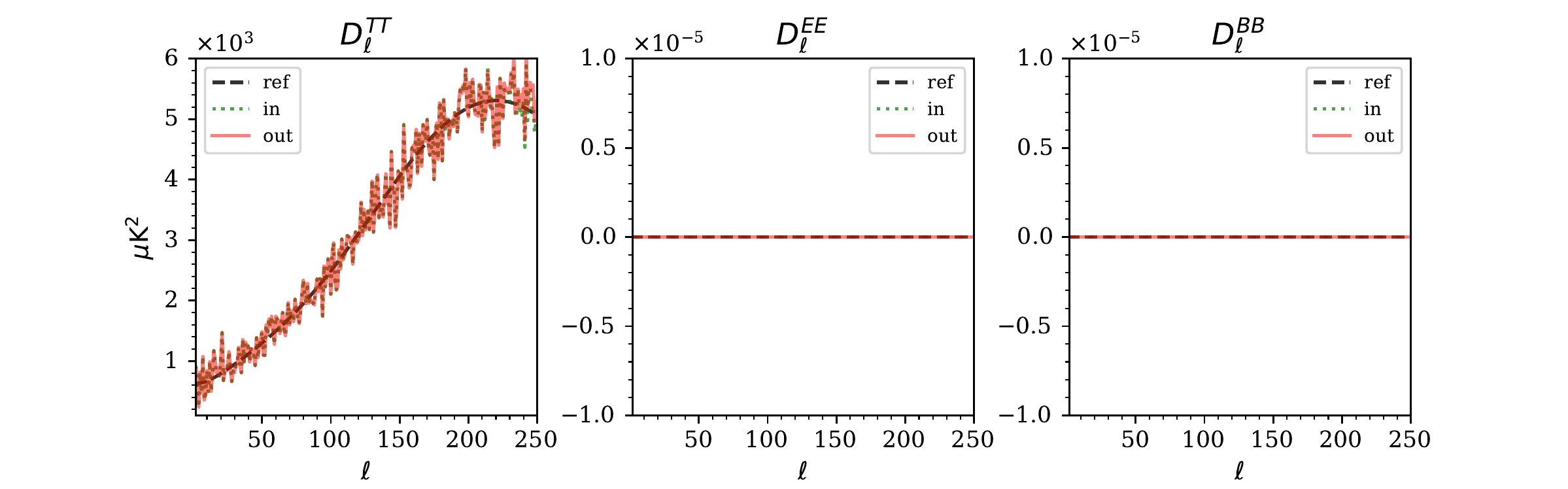}
	\includegraphics[width=1\textwidth, trim = {1.2cm 0.0cm 1.2cm 0.0cm}, clip ]{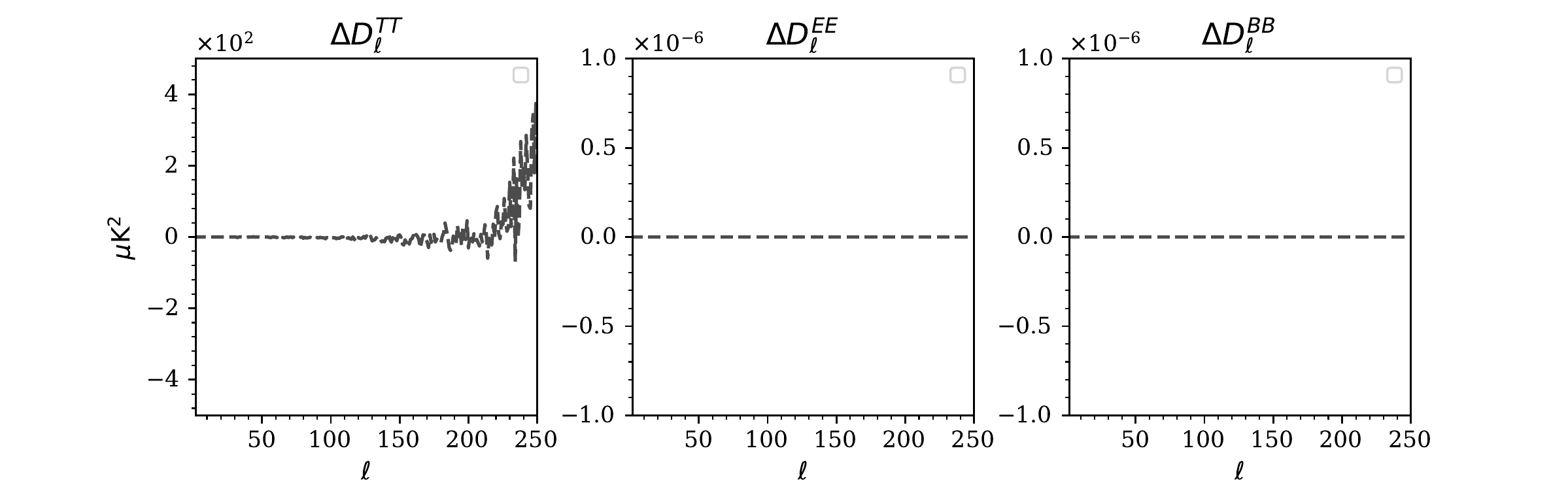}
	\caption{Upper plot: power spectra from PISCO generated TOD (``out'' label) for the whole sky CMB test with random pointing offsets from pixel centers on an unpolarized sky. $D_\ell$ used to generate the input maps are the black, dashed curve (``ref'' label). The input sky corresponds to a CMB with $r=0.0$, and its power spectra are plotted in dotted green (``in'' label). The scale of the y-axis is $\mu\rm{K}^2$ for all three plots. Bottom: difference between power spectra of PISCO generated TOD and the power spectra of the input map. These plots show no signs of temperature to polarization leakage. }
	\label{fig::pisco4wholesky_random_offsets_unpol}
\end{figure}

Results for this test are summarized in Figure \ref{fig::pisco4wholesky_random_offsets}. We observe that the discrepancy between power spectra becomes significant when random offsets are introduced into the pointing. One notorious effect is the appearance of a spurious B-mode signal. To explore the origin of this artifact, we ran one more similar simulation using an unpolarized CMB as input. These results (see Figure \ref{fig::pisco4wholesky_random_offsets_unpol}) indicate that the effect on the B-mode spectrum that we see in Figure \ref{fig::pisco4wholesky_random_offsets} is E-mode to B-mode leakage. We believe the origin of this leakage is related to an uncompensated coupling of intra-pixel gradients into the map-making process. A more detailed description of this phenomena was given in \cite{2005poutanen}. 

\begin{figure}
	\centering
	\includegraphics[width=1\textwidth, trim = {1.2cm 0.0cm 1.2cm 0.0cm}, clip ]{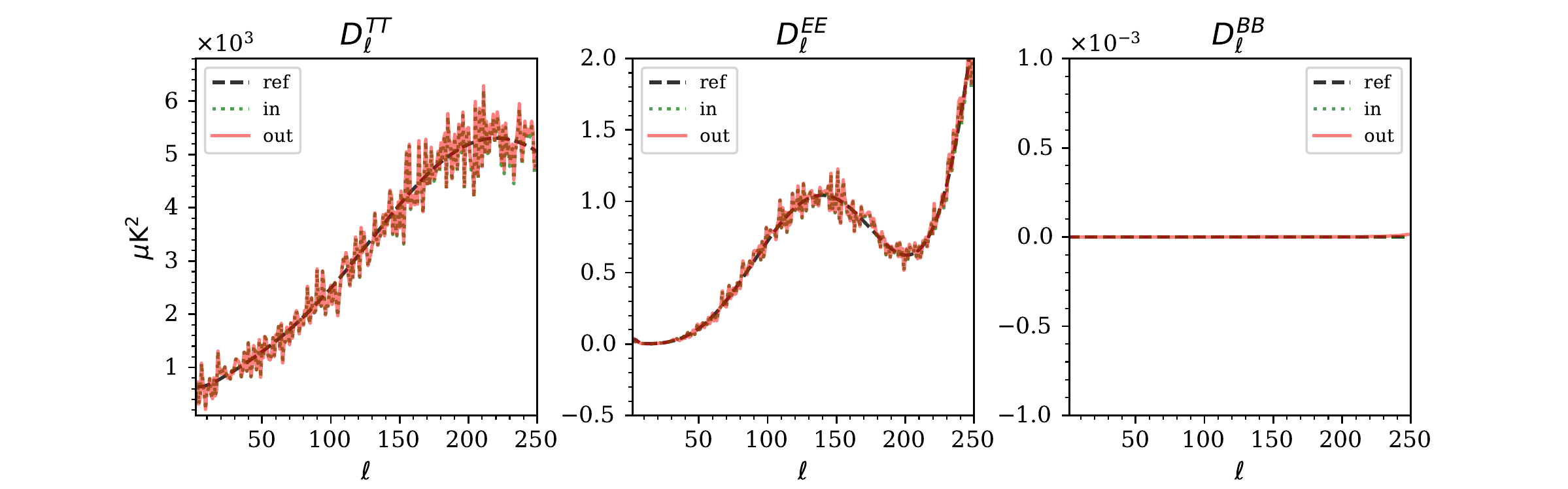}
	\includegraphics[width=1\textwidth, trim = {1.2cm 0.0cm 1.2cm 0.0cm}, clip ]{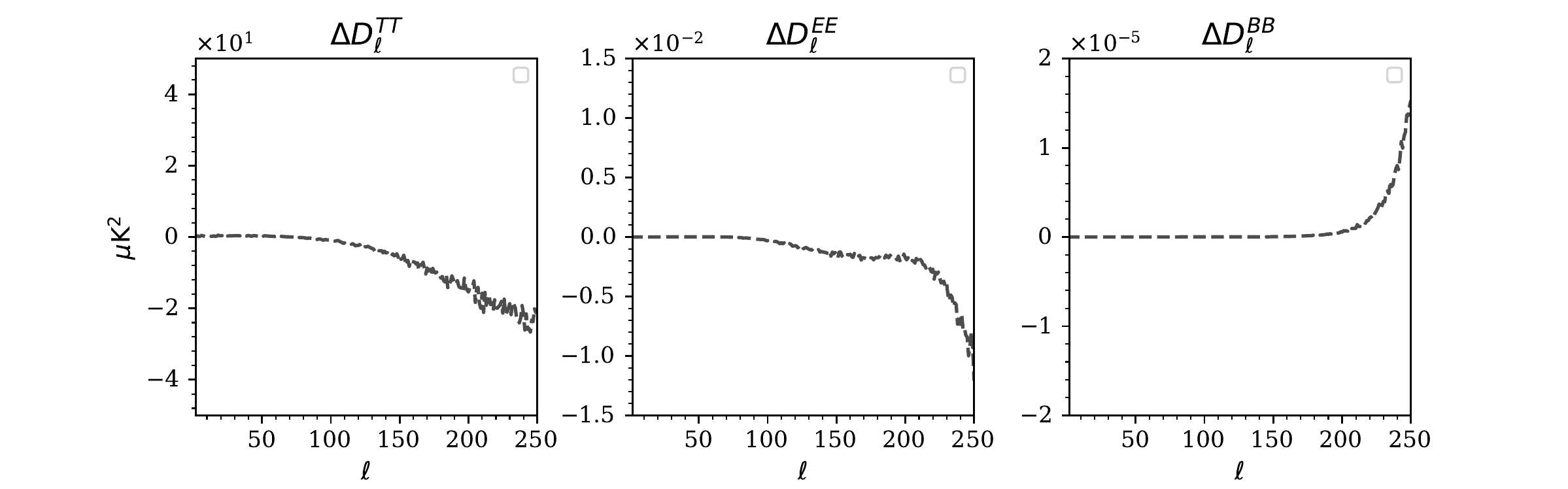}
	\caption{Upper plot: power spectra from PISCO generated TOD (``out'' label) for the whole sky CMB test with random pointing offsets symmetric about pixel centers. $D_\ell$ used to generate the input maps are the black, dashed curve (``ref'' label). The input sky corresponds to a CMB with $r=0.0$, and its power spectra are plotted in dotted green (``in'' label). The scale of the y-axis is $\mu\rm{K}^2$ for all three plots. Bottom: difference between power spectra of PISCO generated TOD and the power spectra of the input map. The spurious B-mode signal reaches $1.5 \times 10^{-5} \, \mu\mathrm{K}^2$ al $\ell=250$. }
	\label{fig::pisco4wholesky_symm_offsets}
\end{figure}

The hypothesis of intra-pixel gradient coupling was validated against a variation of the simulation described at the beginning of this section. The main difference was that random pointing offsets were 
symmetric with respect to pixel centers; for each random offset, an equal and opposite offset was introduced. The main driver for this test was the idea that, at the angular scale of an output map pixel (roughly 0.45 degrees for NSIDE 128), gradients on the sky are are likely to be linear to first order. This would imply that adding an opposite offset to the pointing stream should, in principle, suppress the observed leakage, as the gradient-coupled signal would approximately cancel out. Results for this simulation (see Figure \ref{fig::pisco4wholesky_symm_offsets}) support this theory, as the spurious B-mode signal is suppressed by around 4 orders of magnitude. It is worth noting that symmetrizing the offsets also brings down the residuals of the temperature and E-mode power spectra to levels comparable to those shown on Figure \ref{fig::pisco4wholesky}.

%The main result of this simulation is shown in Figure \ref{fig::pisco4wholesky}. A single simulation took less than 5 seconds on the testing machine. We report no indications of leakage from temperature to polarization, nor from E-modes to B-modes. It is worth noting that this process was repeated using several realizations of roughly the same cosmology, but with varying values of $r$. We found no evidence of leakage in any of these cases. It appears to be a tendency on the residuals to become larger at smaller angular scales. 

\section{A more realistic CMB experiment}
\label{sec::realistic_cmb_experiment}

The tests performed in \S\ref{subsec::ideal_full_sky} correspond to an idealized case. In this section, we show results for a simple, but more realistic, application of PISCO. The focus of this section is on demonstrating that the implementation of the model presented in \S\ref{sec::antennas} and \S\ref{sec::pisco} is able to reproduce the expected effect of pointing and beam mismatch, which is T to P leakage (see \cite{2007MNRAS.376.1767O}), and uneven intra-pixel coverage, which is P to P leakage (see \cite{2005poutanen}). As in the previous section, we did not include other systematic effects like gain imbalance, detector efficiency, cross-polarization responses and far sidelobes.
 
\subsection{Description of the simulations}

Since designing such an experiment from scratch is a challenging task, we turned to simulating an ongoing mission, the Cosmology Large Angular Scale Surveyor, CLASS. CLASS aims at characterizing the CMB anisotropy field at large angular scales, particularly the power spectra of B-modes and E-modes, looking for evidence of inflation. While the experiment is composed of 4 telescopes, the simulation focuses on the one observing at the lowest frequency (38 GHz), the Q-band receiver, which has a 1.5 degree beam. This decision was taken for computational reasons, mainly because Q-band has lower detector count compared to higher frequency receivers. We note that CLASS uses modulation techniques to increase its sensitivity to CMB polarization. The effects of modulation related systematics have been described elsewhere (\cite{2016ApJ...818..151M}). For a more detailed description of CLASS, the reader is referred to \cite{2016SPIE.9914E..1KH}, \cite{2014SPIE.9153E..1IE} and \cite{2019ApJ...876..126A}.

\subsubsection{Sky model}

To generate the maps for this simulation, we followed a procedure similar to the one described in \S\ref{subsec::sky_model}. The main difference is the substitution of a temperature only CMB for the first two tests, which was used to check for T to P leakage caused by effects that were not present in the tests presented in \S\ref{sec::validation}. In addition, all maps use the HEALPix pixelization with an NSIDE parameter of 128 for both sky and output.

\subsubsection{Pointing}

The scanning strategy of CLASS consists of constant elevation scans (CES). Elevation is kept at $45^{\circ}$ while the telescope rotates $720^\circ$ in azimuth at $1$ degree per second. Under normal survey conditions, the telescopes scan nearly 24 hours a day. The boresight is rotated from $-45^{\circ}$ to $+45^{\circ}$ by $15^{\circ}$ per day on a weekly schedule. This scanning strategy, in combination with the large CLASS field of view results in the telescopes covering more than 70$\%$ of the sky. In addition, because of the boresight rotation, only seven days are needed to provide excellent position angle coverage. While CLASS records data $200$ times per second, the pointing streams were generated at $20$ Hz. Down-sampling by a factor of 10 results in a ten-fold decrease in computation time with the median number of hits per pixel still being on the order of thousands. Even with this significant reduction, the pointing stream contained more than 870 million samples.

Equatorial coordinates of every detector were computed from the horizontal coordinates produced by the scanning strategy and the beam center offsets of every detector from the center of the array.
Representative beam center offsets for the Q-band receiver were provided by the CLASS collaboration. Two streams were generated by considering detector pairs to have matched or mismatched offsets. The case of matched offsets was simulated by forcing each pair to share the same beam center offsets, which in turn were calculated as the average of the individual pair offsets. 

\subsubsection{Beamsor model}

The CLASS collaboration also provided us with representative main beam parameters. The main beam parameters correspond to FWHM in the East-West direction ($\rm{FWHM}_x$), FWHM in the North-South direction ($\rm{FWHM}_y$) and the rotation angle of the major axis of the corresponding elliptical profile. The simplicity of the beams allows a further speed-up in the computation by restricting the convolution to a $5^\circ$ disc around the beam center. This value was chosen as, for a unit normalized Gaussian beam, a pixel that is $5^\circ$ away from the centroid of the $1.5^{\circ}$ FWHM beam of the CLASS Q band receiver has a value of $\approx 10^{-14}$, roughly the limit of double precision arithmetic. All beamsors were pixelated using $\rm{NSIDE}=512$. The finite resolution of the beamsor produces an error in the amplitude of simulated TOD that is less than $0.1\%$ compared to analytical estimations, which is consistent with the results described in \S\ref{sec::validation}.

\subsubsection{Power spectra and beam transfer function}

Given that CLASS only covers $\approx 70\%$ of the sky, computing the power spectra of simulated maps requires the use of a tool than can handle a sky mask. For this reason, we changed the estimator from \texttt{anafast} to \texttt{Spice} through the \texttt{PolSpice} implementation (see \cite{2004MNRAS.350..914C}). In addition, the CLASS collaboration provided a realistic sky mask that includes the galactic plane as well as the sky outside the survey boundaries.

While the effect that smooth, circularly symmetric beams have on the power spectra is known (see \cite{2003ApJS..148...39P}), beams for CLASS have non-zero eccentricity and so we need to account for the effect of this on the CMB power spectra. This is done by computing the radial profile of each beam through an analytical integration over $\phi$, averaging the radial profiles of all of the beams in the array receiver and then calculating the harmonic transform of the average profile to use as the beam transfer function in deconvolving the power spectra.

In addition to the beam transfer function, the power spectra of the maps produced by PISCO were compensated by the pixel window function of a \texttt{HEALPix} pixel of resolution parameter \texttt{NSIDE} $=128$. The pixel window function was computed using the \texttt{pixwin} routine of the \texttt{healpy} package.

\subsubsection{A unified metric for describing mismatch }
\label{sec::mismatch_metric}

\begin{figure}
	\centering
	\includegraphics[width=0.7\textwidth]{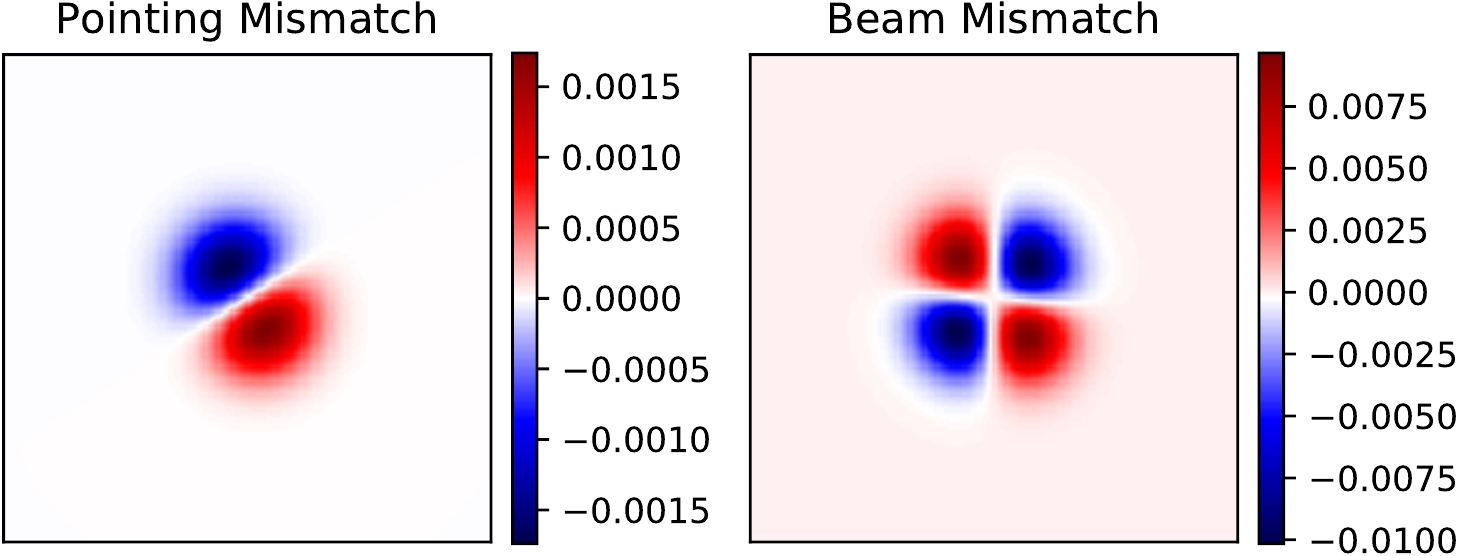}
	\caption{Example pair differenced elliptical Gaussian beams for pointing mismatch and beam mismatch. The resulting metrics are $\delta_p = 0.17\%$ and $\delta_b = 1.0\%$.}
	\label{fig::mismatch_metric}
\end{figure}

While it is simple to use the difference in pointing between two paired detectors to describe pointing mismatch, elliptical beam mismatch involves three quantities: the difference in the semi-major beam widths, the difference in the semi-minor beam widths and the difference in orientation of the two elliptical beams on the sky. When multiple pairs are considered, it becomes unclear how to average these parameters in order to gain insight regarding the degree of mismatch in the set. For this reason, it is desirable to define a metric that can be suitably applied to more than one pair. One such metric can be derived from the difference $\Delta \mathit{b}(\rho,\sigma)$ in the fitted elliptical Gaussian beams for the detector pair of each PSB by considering the maximum difference expressed as a percentage of the normalized beams, namely

\begin{equation}
\delta_{(p,b)} = \rm{max}\left| \Delta \mathit{b}_{(p,b)}(\rho,\sigma) \right|
\end{equation}

\noindent
where the subscripts $p$ and $b$ represent the cases of pointing mismatch and beam mismatch respectively (see Figure \ref{fig::mismatch_metric}). This unified metric is directly proportional to the difference in convolution weighting for each detector pair and can be meaningfully averaged over multiple pairs. 

\subsection{Results and discussion}

\subsubsection{Matched pointing and beams}

For the purpose of comparison, a test was run using matched beams. This was done by averaging both the pointing offsets and elliptical beam parameters for each detector pair in the array. The results of this test are shown in the upper plot of Figure \ref{fig::pisco4class_pointingmismatch}. In this and all subsequent plots the theoretical E-mode and B-mode spectra for a universe with $r=0.1$ were overlaid to provide context. As expected, the result of this test is that there is no T to P leakage present.   

\subsubsection{Pointing mismatch}

CMB experiments that rely on detector pairs are subject to leakage caused by pointing mismatch. As described in the work of \cite{2007MNRAS.376.1767O}, differential pointing between individual detectors belonging to the same PSB couples to gradients in the temperature field.

This test was performed by using the pointing offsets of the individual detectors while averaging the elliptical beam parameters for each detector pair in the array. Applying the metric described in \S\ref{sec::mismatch_metric} to the beams used in this simulation yielded an average value of $\delta_p = 0.42\%$ for the resulting pair differenced pointing mismatch. 

\begin{figure}
    \centering
    \includegraphics[width=1.0\textwidth, trim = {1.5cm 0.0cm 1.5cm 0.0cm}, clip ]{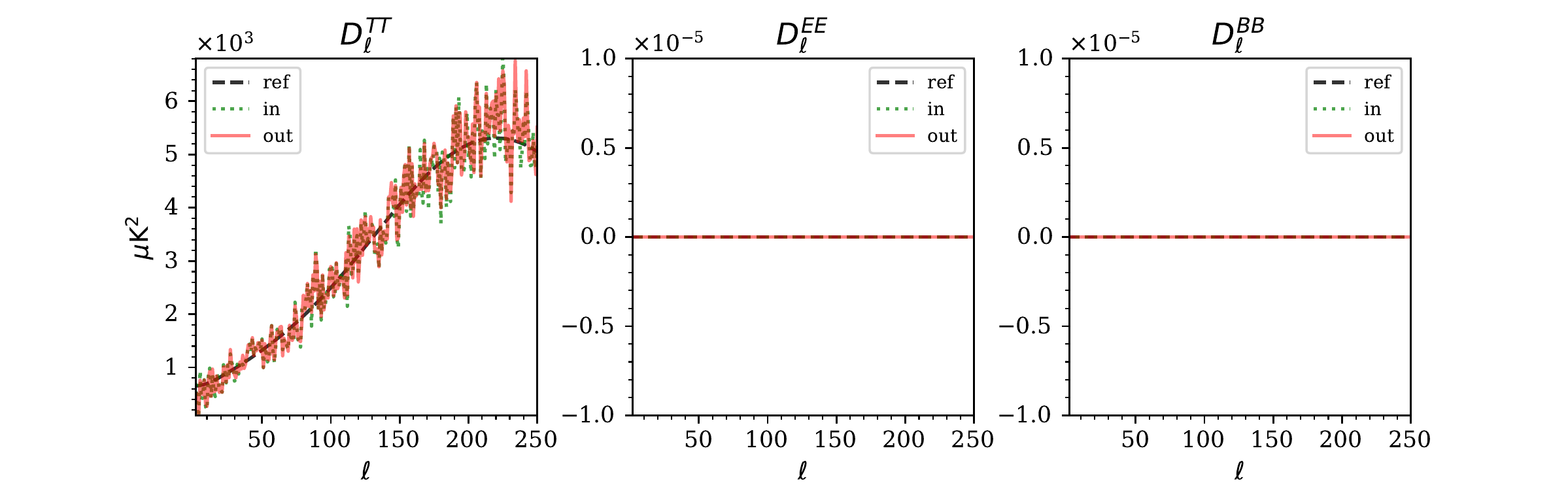}
    \includegraphics[width=1.0\textwidth, trim = {1.5cm 0.0cm 1.5cm 0.0cm}, clip ]{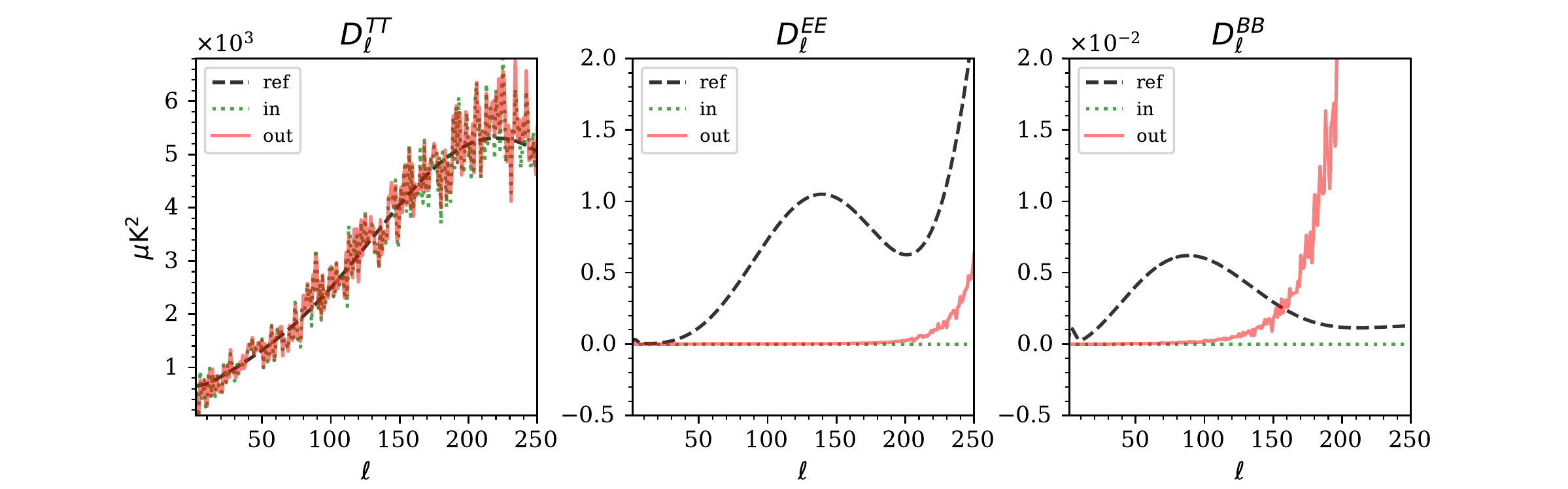}
	\caption{Resulting power spectra for a realistic simulation using matched pointing (upper figure), and mismatched pointing (bottom figure). The input CMB was unpolarized, but reference E-mode and B-mode power spectra for a CMB with $r=0.1$ are shown using the black, dashed line. Leakage from $TT$ to $EE$ and $BB$ power spectra is present for the simulations with mismatched pointing. The spurious leakage signal becomes dominant compared to reference B-modes about $\ell \approx 160$, reaching $0.02\, \mu\mathrm{K}^2$ at $\ell = 190$.}
	\label{fig::pisco4class_pointingmismatch}
\end{figure}

The leakage caused as a result of this pointing mismatch is presented in the lower plot of Figure \ref{fig::pisco4class_pointingmismatch}. We found that the spurious B-mode signal produced by pointing mismatch reaches $0.02\, \mu\mathrm{K}^2$ at $\ell = 190$, and becomes dominant with respect to a reference B-mode spectra with $r=0.1$ at $\ell=160$.

\subsubsection{Beam mismatch}

\begin{figure}

	\centering
	\includegraphics[width=1.0\textwidth, trim = {1.5cm 0.0cm 1.5cm 0.0cm}, clip ]{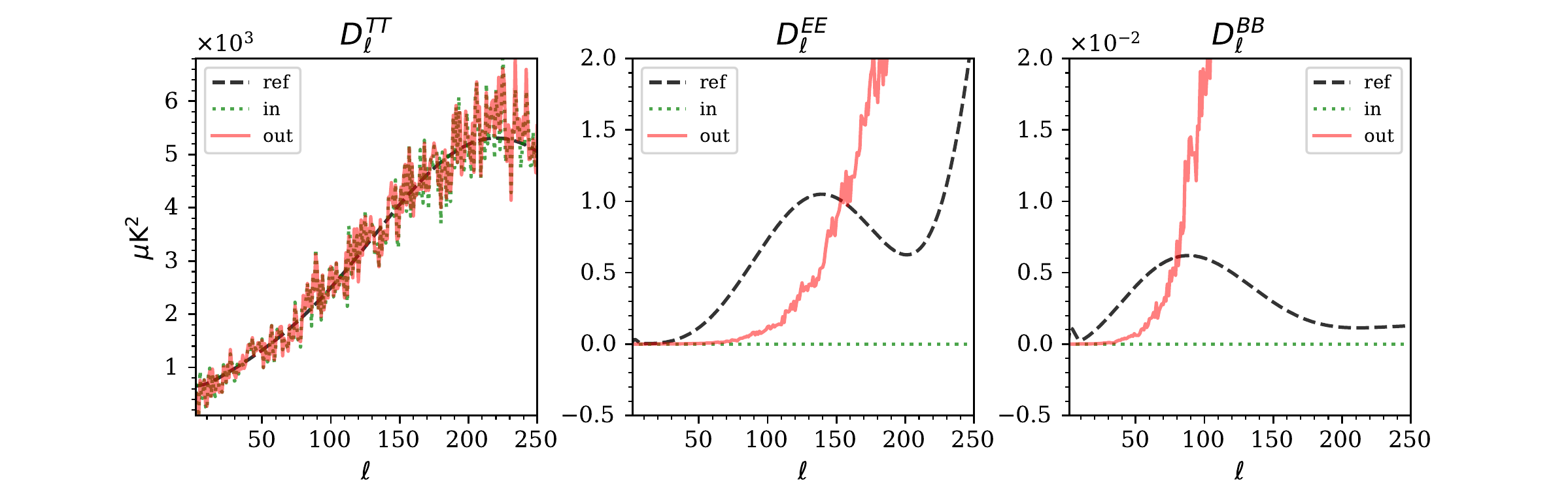}
	\caption{Resulting power spectra of a realistic simulation showing the effects of beam mismatch. The effects of pointing mismatch were removed from this simulation. The input CMB was unpolarized, so that the resulting E-mode and B-mode power spectra can only be a result of T to P leakage. Reference E-mode and B-mode power spectra for a CMB with $r=0.1$ are shown using the black, dashed line. The amplitude of the B-mode leakage signal reaches roughly $0.02\, \mu\rm{K}^2$ at $\ell=100$ and becomes dominant when compared to a B-mode power spectrum (for $r=0.1$) above $\ell = 80$. Similarly, the E-mode power spectrum caused by leakage reaches around $1 \, \mu\rm{K}^2$ at $\ell=150$, and dominates over the cosmological signal at smaller angular scales.}
	\label{fig::pisco4class_beammismatch}
\end{figure}

Beam mismatch produces leakage from temperature to polarization by introducing a coupling between the beam and local quadropole-like patterns in the temperature fields (see \cite{2007MNRAS.376.1767O}) 

This test was performed by using the elliptical beam parameters of the individual detectors while averaging the pointing offsets for each detector pair in the array. Applying the metric described in \S\ref{sec::mismatch_metric} to the beams used in this simulation yielded an average value of  $\delta_b = 1.73\%$ for the resulting pair differenced beam mismatch. 

Figure \ref{fig::pisco4class_beammismatch} shows the resulting power spectra. The leakage reaches around $0.02\, \mu\rm{K}^2$ at $\ell=100$, and the spurious B-mode signal becomes dominant when compared to the theoretical spectrum of primordial B-modes for $r=0.1$ after $\ell=80$. We note that leakage also becomes dominant with respect to the E-mode power spectrum above $\ell=150$. 

\subsubsection{Uneven intra-pixel coverage}

\begin{figure}
	\centering
	\includegraphics[width=1.0\textwidth, trim = {1.5cm 0.0cm 1.5cm 0.0cm}, clip ]{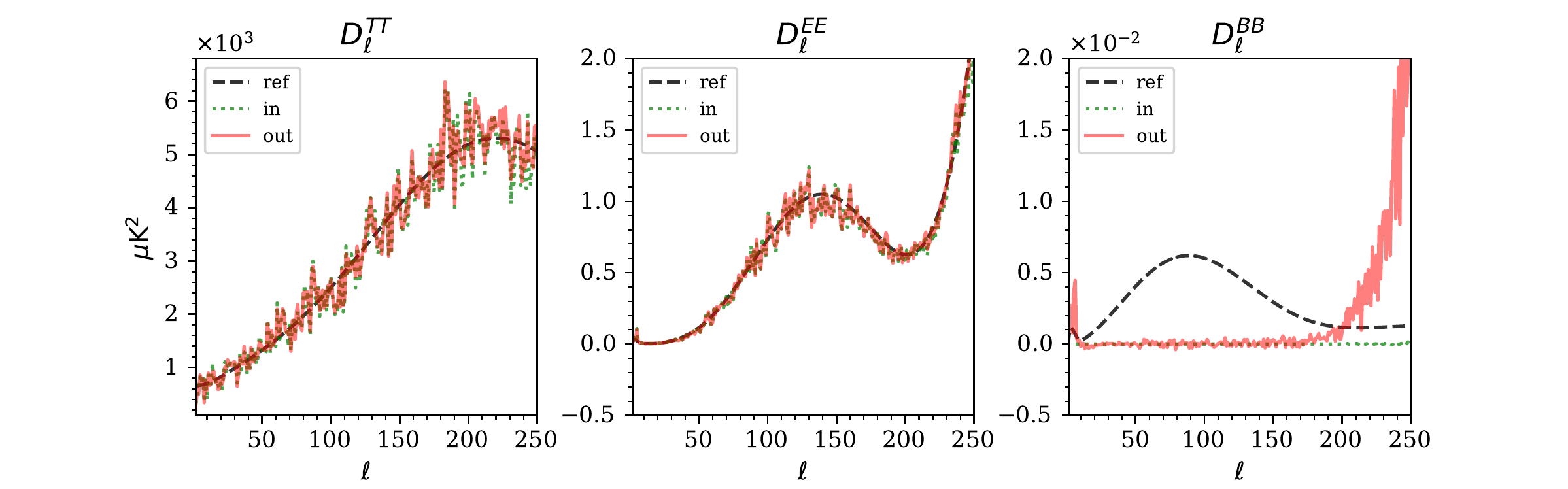}
	\caption{Resulting power spectra for the case of matched pointing with uneven intra-pixel coverage using as input a polarized CMB without B-modes. The B-mode power spectrum shows non-negligible amounts of a spurious signal not present in the input power B-mode power spectrum. Reference E-mode and B-mode power spectra for a CMB with $r=0.1$ are shown using the purple, dashed line. The spurious B-mode spectrum reaches on the order of $0.02\, \mu \rm{K}^2$ at $\ell = 250$, and becomes dominant compared to primordial B-modes above $\ell=200$.}
	\label{fig::pisco4class_intrapixel}
\end{figure}

As discussed in \S\ref{subsec::ideal_full_sky_offsets}, simulating a CMB experiment using a more realistic scanning strategy can produce another systematic effect at the power spectra level related to the intra-pixel coverage of the sky. The goal of this simulation is to show results of this effect using the pointing of a real experiment.
%In \S\ref{subsec::ideal_full_sky}, all pixels were observed exactly at their centers while, in a real experiment, every sample of the TOD ``hits'' a given pixel at an arbitrary location within it. If the distribution of hits inside a pixel is symmetric with respect to pixel center coordinates, map-making will average all observations and the power spectra from the resulting map will not be affected. However, if this distribution is asymmetric and gradients between pixels are present (recall that pixel space convolution is affected by neighbor pixels) the resulting power spectra may suffer from P to P leakage. This was discussed in more detail in more detail in the literature (see \cite{2005poutanen}) and shown to be a realistic effect in Section \ref{subsec::ideal_full_sky_offsets}.

Figure \ref{fig::pisco4class_intrapixel} shows the result of running the realistic multi-beam simulation with matched pointing and matched beams with the addition of a polarized CMB with $r=0.1$. A nonzero, spurious signal is present in the B-mode power spectrum, becoming dominant over the primordial B-modes around $\ell = 200$. In the upper plot of Figure \ref{fig::pisco4class_pointingmismatch}, the same simulation was performed but with an unpolarized CMB as input, the resulting polarized power spectra being consistent with zero at all scales. This indicates that the effect of uneven intra-pixel coverage in this test is P to P leakage, just as shown in the test performed in \S\ref{subsec::ideal_full_sky_offsets}. This systematic effect is subdominant with respect to the T to P leakage caused by pointing and beam mismatch at all scales, as well as having a lower amplitude than a theoretical primordial B-mode power spectrum for $r=0.1$ up to $\ell=200$. Since this P to P leakage is due to an uneven convolution of the pixel window, it is worth noting that, unlike the other systematic effects presented above, this effect can be mitigated as more data is added when using a well planned scanning strategy.

\section{Conclusions}
\label{sec::conclusions}

In this work, we have presented a method for simulating the interaction between the electromagnetic properties of an antenna and a polarized sky in the context of CMB experiments. We have also described PISCO, a new computer simulation code that implements this method. PISCO is capable of generating beam-convolved timestreams for arbitrary beams, sky models and scanning strategies and is designed to exploit the natural parallelism in the pixel space convolution algorithm by offloading the computation to the GPU. We performed tests applying PISCO to several scenarios: point source convolution, an ideal and a more realistic CMB experiment, this last set of tests was based on the CLASS experiment. We showed that PISCO is able to reproduce the expected effects that pointing mismatch, beam mismatch and uneven intra-pixel coverage have on CMB power spectra. 

We are not aware of any other publicly available code that performs pixel space convolution using a formalism that is similar to one used by PISCO, the only documented code that performs pixel space convolution being FEBeCop. We believe PISCO will become a valuable tool when modeling the effect of noise sources that are difficult to include in harmonic space implementations like the one described in \cite{2018arXiv180905034D}. These sources include, but are not limited to, atmospheric turbulence, transient events on the sky and time-dependent behavior of the optics. It is also worth noting that PISCO can naturally include modulation techniques, like half-wave plates, when provided with a modulated model of the beamsor. Finally, the parallelism of the pixel space convolution algorithm provides an opportunity for improvement over the current implementation by harvesting the computational power of modern HPC facilities, particularly those including multiple GPU nodes.

\acknowledgments

The authors thank the CLASS collaboration, the team at Computational Cosmology Center ($\rm{C}^3$) at LBNL, Lo\"ic Maurin and Tobias Marriage for their comments and suggestions. PF and RD thank CONICYT for grants FONDECYT Regular 1141113, Basal AFB-170002, PIA Anillo ACT-1417 and QUIMAL 160009.

\bibliographystyle{plain}

\appendix

\section{Computation of antenna basis coordinates and the co-polarization angle from sky coordinates}

\begin{figure}
	\centering
	\includegraphics[width=0.8\linewidth]{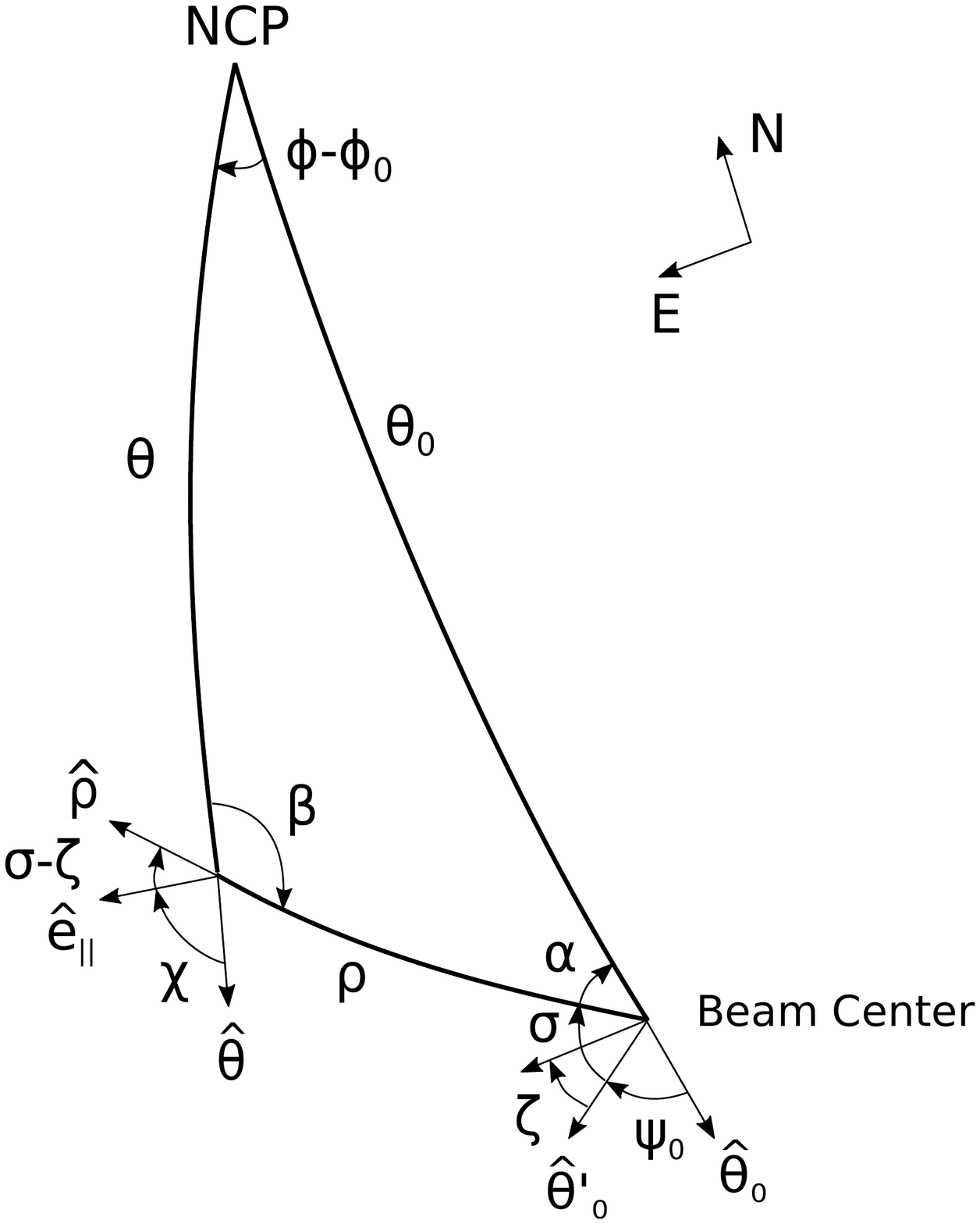}
	\caption{Sky and antenna basis coordinates for beam center pointing and off beam center pointing from the viewpoint of an observer looking at the sky. Here $\mathrm{\texttt{NCP}}$ is the 
		     North Celestial Pole and $\chi$ is the angle between $\hat{e}_{\co}$ and $\hat{\theta}$ according to Ludwig's 3rd definition. }
	\label{fig::figure10}
\end{figure}
For a beam center pointing $(\theta_0, \phi_0)$, rotation angle $\psi_0$ and off beam center pointing $(\theta, \phi)$ in the sky basis, we can derive the antenna basis coordinates $(\rho,\sigma)$ and the angle $\chi$ between $\hat{e}_{\co}$ and $+Q$ ($\hat{\theta}$) by using spherical trigonometry (see Figure \ref{fig::figure10}). The identities used in this derivation are: the law of cosines, the law of sines and the analogue (or five part) formula.

Defining $\Delta \phi \equiv \phi - \phi_0$, the antenna basis coordinate $\rho$ is given by
\begin{align}
\rho  &= \arccos( \cos(\theta) \cos(\theta_0) + \sin(\theta) \sin(\theta_0) \cos(\Delta \phi) ).
\end{align}
Then, defining $\alpha$ as the angle between $\rho$ and $\theta_0$, the antenna basis coordinate $\sigma$ is given by
\begin{align}
\sin(\alpha) &= \frac{\sin(\theta) \sin(\Delta \phi)}{\sin(\rho)} & \\
\cos(\alpha) &= \frac{\cos(\theta) \sin(\theta_0) - \sin(\theta) \cos(\theta_0) \cos(\Delta \phi)}{\sin(\rho)} & \\ 
\alpha       &=  \arctan \left(\frac{\sin(\theta) \sin(\Delta \phi)}{\cos(\theta) \sin(\theta_0) - \sin(\theta) \cos(\theta_0) \cos(\Delta \phi)} \right)  & \\
\sigma &= 180^{\circ} - \psi_0 - \alpha. &
\end{align}
Following Ludwig's third definition of cross polarization, substituting $\sigma + 90^{\circ}$ for Ludwig's $\phi$ (see \cite{1140406}), the unit vectors of co and cross polarization for a detector aligned at an angle $\zeta$ with respect to the basis vector $\hat{\theta}'_0$ in the antenna basis are
\begin{align}
\hat{e}_{\co} &=  \quad \cos(\sigma - \zeta) \hat{\rho} - \sin(\sigma - \zeta) \hat{\sigma} & \\
\hat{e}_{\cx} &=      - \sin(\sigma - \zeta) \hat{\rho} - \cos(\sigma - \zeta) \hat{\sigma}. & 
\end{align}
Thus $\hat{e}_{\co}$ is offset from $\hat{\rho}$ by the angle $\zeta - \sigma$. Then, defining $\beta$ as the angle between $\theta$ and $\rho$, yields
\begin{align}
\sin(\beta) &= \frac{\sin(\theta_0) \sin(\Delta \phi)}{\sin(\rho)} & \\
\cos(\beta) &= \frac{\cos(\theta_0) \sin(\theta) - \sin(\theta_0) \cos(\theta) \cos(\Delta \phi)}{\sin(\rho)} & \\ 
\beta       &=  \arctan \left(\frac{\sin(\theta_0) \sin(\Delta \phi)}{\cos(\theta_0) \sin(\theta) - \sin(\theta_0) \cos(\theta) \cos(\Delta \phi)} \right)  & \\
\chi &= \beta + \zeta - \sigma. &
\end{align}

\end{document}